\documentclass[orivec]{llncs} 
\usepackage[T1]{fontenc}
\usepackage[utf8]{inputenc}
\usepackage[english]{babel}
\usepackage{stmaryrd}
\usepackage{amssymb}
\usepackage{amsmath}
\usepackage{hyperref}
\usepackage[matrix,arrow,curve]{xy}
\CompileMatrices
\usepackage{lmodern}
\usepackage{listings}
\usepackage{tikz}
\usepackage{wrapfig}
\usepackage{setspace}
\usepackage{times}
\usepackage[small,compact]{titlesec}

\newcommand{\comment}[1]{}

\newcommand{\gramor}{\quad|\quad}

\newcommand{\resources}{\mathcal{R}}
\newcommand\set[1]{\{#1\}}
\newcommand\setof[2]{\set{#1\ /\ #2}}
\newcommand{\R}{\mathbb{R}}
\newcommand{\fatone}{\bf 1}
\newcommand{\N}{\mathbb{N}}
\newcommand{\Nint}[2]{[#1:#2[}
\newcommand{\C}{\mathcal{C}}
\newcommand{\nbd}{\nobreakdash-\hspace{0pt}}
\newcommand{\ie}{i.e.~}
\newcommand{\cf}{cf.~}

\newcommand{\resp}{resp.~}
\newcommand{\pone}{\mathbf{1}}
\renewcommand{\P}[1]{P_{#1}}
\newcommand{\V}[1]{V_{#1}}
\newcommand{\ce}{\mathop{\mathrm{e}}}
\newcommand{\ci}{\mathrm{i}}

\newcommand{\qeq}{\quad=\quad}
\newcommand{\dui}{\vec{I}} 

\newcommand{\Z}{\mathbb{Z}}
\newcommand{\vxym}[1]{\vcenter{\xymatrix{#1}}}
\newcommand{\capacity}[1]{\kappa_{#1}}
\newcommand{\tspace}[1]{\vec T(#1)}

\newcommand{\Mat}{\mathcal{M}}

\newcommand{\transition}[1]{\overset{#1}\to}
\newcommand{\transitionpath}[1]{\overset{#1}\twoheadrightarrow}
\newcommand{\lbl}[1]{\ell(#1)}
\newcommand{\cone}{\text{\normalsize\textcircled{\scriptsize 1}}}
\newcommand{\ctwo}{\text{\normalsize\textcircled{\scriptsize 2}}}
\newcommand{\qtand}{\quad\text{and}\quad}
\newcommand{\D}{\mathcal{D}}
\newcommand{\glue}[1]{\oplus_{#1}}

\renewcommand{\leq}{\leqslant}
\renewcommand{\geq}{\geqslant}
\newcommand{\pa}[1]{\left(#1\right)}

\title{Trace Spaces: an Efficient New Technique for State-Space Reduction}
\author{L. Fajstrup\inst{1}\and É. Goubault\inst{2}\and E. Haucourt\inst{2}\and S. Mimram\inst{2}\and M. Raussen\inst{1}}
\institute{Department of Mathematical Sciences, Aalborg University\and CEA,
  LIST\thanks{This work has been supported by the PANDA (``Parallel and
    Distributed Analysis'', ANR-09-BLAN-0169) French ANR project and by ESF project ACAT.}}

\hypersetup{
  pdftitle={\csname @title\endcsname},
  pdfauthor={L. Fajstrup, É. Goubault, E. Haucourt, S. Mimram, M. Raussen},
  unicode=true,
  colorlinks=true,
  linkcolor=black,
  citecolor=black,
  urlcolor=black
}

\renewcommand{\C}{\mathcal{C}}

\begin{document}


\maketitle

\begin{abstract}
  State-space reduction techniques, used primarily in model-checkers, all rely
  on the idea that some actions are independent, hence could be taken in any
  (respective) order while put in parallel, without changing the semantics. It
  is thus not necessary to consider all execution paths in the interleaving
  semantics of a concurrent program, but rather some equivalence classes. The
  purpose of this paper is to describe a new algorithm to compute such
  equivalence classes, and a representative per class, which is based on ideas
  originating in algebraic topology. We introduce a geometric semantics of
  concurrent languages, where programs are interpreted as directed topological
  spaces, and study its properties in order to devise an algorithm for computing
  dihomotopy classes of execution paths. In particular, our algorithm is able to
  compute a control-flow graph for concurrent programs, possibly containing
  loops, which is ``as reduced as possible'' in the sense that it generates
  traces modulo equivalence. A preliminary implementation was achieved, showing
  promising results towards efficient methods to analyze concurrent programs,
  with very promising results compared to partial-order reduction techniques.
\end{abstract}

\section*{Introduction}

Formal verification of concurrent programs is traditionally considered as a
difficult problem because it might involve checking all their possible
schedulings, in order to verify all the behaviors the programs may exhibit. This
is particularly the case for checking for liveness or reachability properties,
or in the case of verification methods that imply traversal of some important
parts of the graph of execution, such as model-checking \cite{modelchecking} and
abstract testing \cite{abstracttesting}. Fortunately, many of the possible
executions are equivalent (we say \emph{dihomotopic}) in the sense that one can
be obtained from the other by permuting independent instructions, therefore
giving rise to the same results. In order to analyze a program, it is thus
enough (and much faster) to analyze one representative in each dihomotopy class
of execution traces.

We introduce in this paper a new algorithm to reduce the state-space explosion
during the analysis of concurrent systems. It is based on former work of some of
the authors, most notably~\cite{raussen2010simplicial} where the notion of trace
space is introduced and studied, and also builds up considerably on the
geometric semantics approach to concurrent systems, as developed
in~\cite{tcs}. Some fundamentals of the mathematics involved can be found
in~\cite{grandis}. The main contributions of this article are the following: we
develop and improve the algorithms for computing trace spaces
of~\cite{raussen2010simplicial} by reformulating them in order to devise an
efficient implementation for them, we generalize this algorithm to programs
which may contain loops and thus exhibit an infinite number of behaviors, we
apply these algorithms to a toy shared-memory language whose semantics is given
in the style of~\cite{fajstrup2000infinitely}, but in this paper, formulated in
terms of d-spaces~\cite{grandis},
and we report on the implementation and experimentation of our algorithms on
trace spaces -- an industrial case-study using those methods is also detailed
in~\cite{rigorous-evidence}.

Stubborn sets~\cite{AVStubborn2}, sleep sets and persistent
sets~\cite{PGPWUsing} are among the most popular methods used for diminishing
the complexity of model-checking using transition systems; they are in
particular used in SPIN~\cite{spin}, with which we compare our work
experimentally in Section~\ref{benchmarks}. They are based on semantic
observations using Petri nets in the first case and Mazurkiewicz trace theory in
the other one. We believe that these are special forms of dihomotopy-based
reduction as developed in this paper when cast in our geometric framework, using
the adjunctions of~\cite{getco2010}. Of course, the trace spaces we are
computing have some acquaintance with traces as found in trace theory
\cite{tracetheory}: basically, traces in trace theory are points of trace
spaces, and composition of traces modulo dihomotopy is concatenation in trace
theory. Trace spaces are more general in that they consider general directed
topological spaces and not just partially commutative monoids; they also include
all information related to higher-dimensional (di-)homotopy categories, and not
just the fundamental category, as in trace theory. Trace spaces are also linked
with component categories, introduced by some of the authors \cite{apcs,apcs2},
and connected components of trace spaces can also be computed using the
algorithm introduced in \cite{concur05}.

\paragraph{Contents of the paper.}
We first define formally the programming language we are considering
(Section~\ref{language}) 
as well as an associated geometric semantics, (Section~\ref{geomsem}). We then
introduce an algorithm for computing an effective combinatorial representation
of trace spaces as well as an efficient implementation of it
(Section~\ref{computing}), and extend this algorithm in order to handle program
containing loops (Section~\ref{programswithloops}). Finally, we discuss various
applications, in particular to static analysis (Section~\ref{sec:static-anal})
and possible extensions of the algorithm and conclude.

\section{Geometric semantics of concurrent processes}

\subsection{A toy shared-memory concurrent language}
\label{language}

In this paper, we consider a toy imperative shared-memory concurrent language as
grounds for experimentation. In this formalism, a program can be constituted of
multiple subprograms which are run in parallel. The environment provides a set
of resources~$\resources$, where each resource~$a\in\resources$ can be used by
at most~$\capacity{a}$ subprograms at the same time, the
integer~$\capacity{a}\in\N$ being called the \emph{capacity} of the
resource~$a$. In particular, a \emph{mutex} is a resource of capacity~$1$.

Whenever a program wants to access a resource~$a$, it should acquire a lock by
performing the action~$\P{a}$ which allows access to~$a$, if the lock is
granted. Once it does not need the resource anymore, the program can release the
lock by performing the action~$\V{a}$, following again the notation set up by
Dijkstra~\cite{DijkstraPV}. If a subprogram tries to acquire a lock on a
resource~$a$ when the resource has already been locked~$\capacity{a}$ times, the
subprogram is stuck until the resource is released by an other subprogram. In
order to be realistic even though simple, the language considered here also
comprises a sequential composition operator~$.$, a non-deterministic choice
operator~$+$ and a loop construct~$(-)^*$, with similar semantics as in regular
languages (it should be thought as a \texttt{while} construct), as well as a
parallel composition operator~$|$ to launch two subprograms in parallel.

Programs~$p$ are defined by the following grammar:
\[
p
\quad::=\quad
\pone
\gramor
\P{a}
\gramor
\V{a}
\gramor
p.p
\gramor
p|p
\gramor
p+p
\gramor
p^*
\]
Programs are considered modulo a \emph{structural congruence}~$\equiv$ which
imposes that operators $.$, $+$ and~$|$ are associative and admit~$\pone$ as
neutral element.
A \emph{thread} is a program which does not contain the parallel composition
operator~$|$.

\subsection{Geometric semantics}
\label{geomsem}

We introduce here a semantics based on
(directed) topological spaces. The geometric semantics will allow a
different representation of~$n$ pairwise independent actions (as the surface of
an $n$-cube) and~$n$ truly concurrent actions as the full $n$-cube.

We denote by~$I=[0,1]\subseteq\R$ the standard euclidean interval. A
\emph{path}~$p$ in a topological space~$X$ is a continuous map~$p:I\to X$, and
the points~$p(0)$ and~$p(1)$ are respectively called the \emph{source} and
\emph{target} of the path. Given two paths~$p$ and~$q$ such that $p(1)=q(0)$, we
define their \emph{concatenation} as the path~$p\cdot q$ defined by
\[
(p\cdot q)(t)=
\begin{cases}
  p(2t)&\text{if $0\leq t\leq 1/2$}\\
  q(2t-1)&\text{if $1/2\leq t\leq 1$}
\end{cases}
\]

A topological space can be equipped with a notion of ``direction'' as
follows~\cite{grandis}:

\begin{definition}
  A \emph{directed topological space} (or \emph{d-space} for short) $X=(X,dX)$
  consists of a topological space~$X$ together with a set~$dX$ of paths in~$X$
  (the \emph{directed paths}) such that
  \begin{enumerate}
  \item \emph{constant paths}: every constant path is directed,
  \item \emph{reparametrization}: $dX$ is closed under precomposition with (non
    necessarily surjective) increasing maps \hbox{$I\to I$}, which are called
    \emph{reparametrizations},
  \item \emph{concatenation}: $dX$ is closed under concatenation.
  \end{enumerate}
  A morphism of d-spaces $f:X\to Y$, a \emph{directed map}, is a continuous
  function \hbox{$f:X\to Y$} which preserves directed paths, in the sense that
  $f(dX)\subseteq dY$.
\end{definition}

The category of d-spaces is complete and cocomplete~\cite{grandis}. This allows
us to abstractly define some constructions on d-spaces, which extend usual
constructions on topological spaces, that we detail here explicitly by
describing the associated directed paths.
\begin{itemize}
\item The \emph{terminal d-space} $\star$ is the space reduced to one
  point.
\item The \emph{cartesian product} $X\times Y$ of two d\nbd{}spaces~$X$ and~$Y$
  has $d(X\times Y)=dX\times dY$.
\item The \emph{disjoint union} $X\uplus Y$ of two d\nbd{}spaces~$X$ and~$Y$ is
  such that $d(X\uplus Y)=dX\uplus dY$.
\item The \emph{amalgamation} $X[x=y]$ of two points $x$ and $y$ in a
  d\nbd{}space $X$ is the d\nbd{}space~$X$ where~$x$ and $y$ have been
  identified, together with the expected set of directed paths.
\item Given a d-space~$X$ and a topological space~$Y\subseteq X$, the
  \emph{subspace}~$Y$ can be canonically equipped with a structure of d-space
  by~$dY=\setof{p\in dX}{p(I)\subseteq Y}$.
\end{itemize}
The geometric semantics of a program is defined using those constructions as
follows:

\begin{definition}
  To every program~$p$, we associate a d-space~$G_p$ together with a pair of
  points $b_p,e_p\in G_p$, respectively called \emph{beginning} and \emph{end},
  and a \emph{resource function} $r_p:\resources\times G_p\to\Z$ which indicates
  the number of locks the program holds at a given point. The definition of
  these is done by induction on the structure of~$p$ as follows:

  \medskip
  \noindent
  \begin{tabular}{l|l}
    $G_\pone=\star$,\quad $b_\pone=\ast$,\quad $e_\pone=\ast$,\quad $r_\pone(a,x)=0$
    \\
    \hline
    $G_{\P{a}}=\dui$,\quad $b_{\P{a}}=0$,\quad $e_{\V{a}}=1$,
    &
    $G_{\V a}=\dui$,\quad $b_{\V a}=0$,\quad $e_{\V a}=1$,
    \\
    $r_{\P a}(b,x)=
    \begin{cases}
      -1&\text{if $b=a$ and $x>0$}\\
      0&\text{if $b\neq a$ or $x=0$}\\
    \end{cases}$
    &
    $r_{\V a}(b,x)=
    \begin{cases}
      1&\text{if $b=a$ and $x=1$}\\
      0&\text{if $b\neq a$ or $x<1$}\\
    \end{cases}$
    \\
    \hline
    $G_{p.q}=(G_p\uplus\dui\uplus G_q)[e_p=0,1=b_q]$,
    &
    $G_{p+q}=(G_p\uplus G_q)[b_p=b_q,e_p=e_q]$,
    \\
    $b_{p.q}=b_p$,\quad $e_{p.q}=e_q$,
    &
    $b_{p+q}=b_p$,\quad $e_{p+q}=e_q$,
    \\
    $r_{p.q}(a,x)=
    \begin{cases}
      r_p(a,x)&\text{if $x\in G_p$}\\
      r_p(a,e_p)+r_q(a,x)&\text{if $x\in G_q$}\\
    \end{cases}$
    &
    $r_{p+q}(a,x)=
    \begin{cases}
      r_p(a,x)&\text{if $x\in G_p$}\\
      r_q(a,x)&\text{if $x\in G_q$}
    \end{cases}
    $
    \\
    \hline
    $G_{p|q}=G_p\times G_q$,
    &
    $G_{p^*}=G_p[b_p=e_p]$,
    \\
    $b_{p|q}=(b_p,b_q)$,\quad $e_{p|q}=(e_p,e_q)$,
    &
    $b_{p^*}=b_p$,\quad $e_{p^*}=b_p$,
    \\
    $r_{p|q}(a,(x,y))=r_p(a,x)+r_q(a,y)$
    &
    $r_{p^*}(a,x)=r_p(a,x)$
  \end{tabular}

  \medskip\noindent
  Given a program~$p$, the \emph{forbidden region} is the d-space~$F_p\subseteq
  G_p$ defined by
  \[
  F_p=\setof{x\in G_p}{\exists a\in\resources,\ \capacity{a}+r_p(a,x)<0\text{\ \ or\ \ }r_p(a,x)>0}
  \]
  The \emph{geometric realization} of a process~$p$, is defined as the
  d\nbd{}space $H_p=G_p\setminus F_p$.
\end{definition}

\noindent
We sometimes write~$0$ and~$\infty$ for the beginning and the end points
respectively of a geometric realization, and say that a path~$p:\dui\to G_p$ is
\emph{total} when it has~$0$ as source and~$\infty$ as target. It is easy to
show that the geometric semantics of a program is well-defined in the sense that
two structurally congruent programs give rise to isomorphic geometric
realizations.


\begin{example}
  The processes
  \[
  \P{a}.\V{a}|\P{a}.\V{a}
  \qquad\qquad
  \P{a}.\P{b}.\V{b}.\V{a}|\P{b}.\P{a}.\V{a}.\V{b}
  \qquad\qquad
  \P{a}.(\V{a}.\P{a})^*|\P a.\V a
  \]
  respectively have the following geometric realizations, which all consist of a
  space with some ``holes'', drawn in gray, induced by the forbidden region:
  \[
  \begin{tikzpicture}[scale=0.3]
    \draw (-.3,-.3) node {$b_p$};
    \draw (5.4,5.4) node {$e_p$};
    \draw[->] (0,0) -- (5,0);
    \draw[->] (0,0) -- (0,5);
    \draw[->] (5,0) -- (5,5);
    \draw[->] (0,5) -- (5,5);
    \fill[fill=gray] (1.5,1.5) -- (1.5,3.5) -- (3.5,3.5) -- (3.5,1.5) -- cycle;
  \end{tikzpicture}
  \qquad\qquad
  \begin{tikzpicture}[scale=0.3]
    \draw (-.3,-.3) node {$b_p$};
    \draw (5.4,5.3) node {$e_p$};
    \draw[->] (0,0) -- (5,0);
    \draw[->] (0,0) -- (0,5);
    \draw[->] (5,0) -- (5,5);
    \draw[->] (0,5) -- (5,5);
    \fill[fill=gray] (1,2) -- (1,3) -- (4,3) -- (4,2) -- cycle;
    \fill[fill=gray] (2,1) -- (2,4) -- (3,4) -- (3,1) -- cycle;
  \end{tikzpicture}
  \qquad\qquad
  \begin{tikzpicture}[scale=0.3]
    \fill[fill=gray] (3.2,1) -- (5.3,0.5) -- (8.3,1.9) -- (6.2,2.5) -- cycle;
    \fill[fill=gray] (7.5,.85) -- (8,.2) -- (6.8,.5) -- cycle;
    \draw (-.3,-.3) node {$b_p$};
    \draw (9,0.9) node {$e_p$};
    \draw[->] (0,0) -- (4,-1);
    \draw plot[smooth] coordinates { (4,-1) (10,2) (12,1) (10,-1) (8,1.1)};
    \draw plot[smooth] coordinates { (0,0) (6,3) (11,1.97) };
    \draw plot[smooth] coordinates { (10,-1) (6,.05)};
    \draw[dotted] (4,2.1) -- (8,1.1);
  \end{tikzpicture}
  \]
  The space in the middle is sometimes called the ``Swiss flag'' because of its
  form and is interesting because it exhibits both a deadlock and an unreachable
  region~\cite{tcs}.
\end{example}

\section{Computing trace spaces}
\label{computing}
\subsection{Trace spaces}
\label{tracespaces}
In topology, two paths~$p$ and~$q$ are often considered as equivalent when~$q$
can be obtained by deforming continuously~$p$ (or vice versa), this equivalence
relation being called \emph{homotopy}. The corresponding variant of this
relation in the case of directed topological spaces is called \emph{dihomotopy}
and is formally defined as follows. In the category of d-spaces, the
object~$\dui$ is \emph{exponentiable}, which means that for every d-space~$Y$,
one can associate a d-space~$Y^{\dui}$ such that there is a natural bijection
between morphisms~$X\times\dui\to Y$ and morphisms~$X\to Y^{\dui}$. The
underlying space of~$Y^{\dui}$ is the set of functions~$\dui\to Y$ with the
compact-open topology (also called uniform convergence topology), and the
directed paths \hbox{$h:\dui\to Y^{\dui}$} are the functions such that~$t\mapsto
h(t)(u)$ is increasing for every~$u\in\dui$. Finally, two paths are said to be
dihomotopic when one can be continuously deformed into the other:

\begin{definition}
  \label{def:dihomotopy}
  The \emph{dihomotopy} is defined as the smallest equivalence relation on paths
  such that two directed paths~$p,q:\dui\to X$ are dihomotopic when there exists
  a directed path~$h:\dui\to X^{\dui}$ with~$p$ as source and~$q$ as target.
\end{definition}

\begin{example}
  In the geometric semantics of the program $\P b.\V b.\P a.\V a\ |\ \P a.\V a$,
  the two paths above the hole are dihomotopic, whereas the path below is not
  dihomotopic to the two others:
   \[
   \begin{tikzpicture}[scale=0.6]
     \draw[line width=0.3mm,color=red] (0,0) .. controls (1,1.5) .. (4,2);
     \draw[line width=0.3mm,color=orange] (0,0) .. controls (1.5,.3) .. (2,1) .. controls (2.3,1.6) .. (4,2);
     \draw[line width=0.3mm,color=blue] (0,0) .. controls (3.9,0.1) .. (4,2);
     \draw (0,0) -- (0,2) -- (4,2) -- (4,0) -- cycle;
     \fill[fill=gray] (2.5,.5) -- (2.5,1.5) -- (3.5,1.5) -- (3.5,.5) -- cycle;
     \foreach \x in {0,1} \draw (-1mm,\x+.5) -- (1mm,\x+.5);
     \foreach \x in {0,1,2,3} \draw (\x+.5,-1mm) -- (\x+.5,1mm);
     \draw (.5,-4mm) node {$\P b$};
     \draw (1.5,-4mm) node {$\V b$};
     \draw (2.5,-4mm) node {$\P a$};
     \draw (3.5,-4mm) node {$\V a$};
     \draw (-4mm,.5) node {$\P a$};
     \draw (-4mm,1.5) node {$\V a$};
   \end{tikzpicture}
   \vspace{-2ex}
   \]
\end{example}

\noindent
The intuition underlying the geometric semantics is that two dihomotopic paths
correspond to execution traces differing by inessential commutations of
instructions, thus giving rise to the same result.

Given two points~$x$ and~$y$ of a d-space~$X$, we write~$X(x,y)$ for the subset
of~$X^{\dui}$ consisting of dipaths from~$x$ to~$y$. A \emph{trace} is the
equivalence class of a path modulo surjective reparametrization, and a
\emph{scheduling} is the equivalence class of a trace modulo dihomotopy. We write
$\tspace{X}(x,y)$ for the \emph{trace space} obtained from~$X(x,y)$ by
identifying paths equivalent up to reparametrization, and simply $\tspace{X}$
for $\tspace{X}(0,\infty)$. In particular, we have
$\tspace{X}(x,y)\neq\emptyset$ if and only if there exists a directed path
in~$X$ going from~$x$ to~$y$.

In this section, we reformulate the algorithm for computing the trace
space~\hbox{$\tspace X$} up to dihomotopy equivalence, originally introduced
in~\cite{raussen2010simplicial}, in order to achieve an efficient implementation
of it. For simplicity, we restrict here to spaces which are geometric
realizations of programs of the form
\begin{equation}
  \label{eq:prog}
  p
  \qeq
  p_0\ |\ p_1\ |\ \ldots\ |\ p_{n-1}
\end{equation}
where the~$p_i$ are built up only from~$\pone$, concatenation, resource locking
and resource unlocking (extending the algorithm to programs which may contain
loops requires significant generalizations which are described in
Section~\ref{programswithloops}). In this case, the geometric realization is of
the form
\vspace{-2ex}
\[
G_p
\qeq
\dui^n\setminus\bigcup_{i=0}^{l-1}R^i
\]
where $\dui^n$ denotes the cartesian product of~$n$ copies of~$\dui$, and each
\hbox{$R^i=\prod_{j=0}^{n-1}\dui^i_j$} is a rectangle. We suppose here that
each~$R^i$ is homothetic to the $n$\nbd{}dimensional open rectangle, \ie each
directed interval~$\dui^i_j$ is of the form~$\dui^i_j=]x^i_j,y^i_j[$, and
generalize this at the end of the section. The restrictions on the form of the
programs are introduced here only to simplify our exposition: programs with
choice can be handled by computing the trace spaces on each branch and program
with loops can be handled by suitably unfolding the loops so that all the
possible behaviors are exhibited (a detailed presentation of this is given in
Section~\ref{programswithloops}, which will enable to handle the full
language). We suppose fixed a program with~$n$ threads and~$l$ forbidden open
rectangles, and consistently use the notations above.

\begin{example}
  \label{ex:geom-rel}
  The geometric realization of the programs
  \[
  \P a.\V a.\P b.\V b|\P a.\V a.\P b.\V b
  \qquad\text{and}\qquad
  \P a.\V a.\P b.\V b|\P b.\V b.\P a.\V a
  \]
  are respectively
  \vspace{-4ex}
  \[
  \begin{tikzpicture}[scale=.4]
    \draw[->] (0,0) -- (5,0);
    \draw[->] (0,0) -- (0,5);
    \draw (5.5,0) node {$t_0$};
    \draw (0,5.5) node {$t_1$};
    \fill[fill=gray] (1,1) -- (2,1) -- (2,2) -- (1,2) -- cycle;
    \fill[fill=gray] (3,3) -- (4,3) -- (4,4) -- (3,4) -- cycle;
    \draw (1.5,1.5) node {$0$};
    \draw (3.5,3.5) node {$1$};
    \foreach \x in {1,2,3,4} \draw (\x,-1mm) -- (\x,1mm);
    \foreach \x in {1,2,3,4} \draw (-1mm,\x) -- (1mm,\x);
    \draw (1,-.6) node {$x^0_0$};
    \draw (2,-.6) node {$y^0_0$};
    \draw (3,-.6) node {$x^1_0$};
    \draw (4,-.6) node {$y^1_0$};
    \draw (-.6,1) node {$x^0_1$};
    \draw (-.6,2) node {$y^0_1$};
    \draw (-.6,3) node {$x^1_1$};
    \draw (-.6,4) node {$y^1_1$};
  \end{tikzpicture}
  \qtand
  \begin{tikzpicture}[scale=.4]
    \draw[->] (0,0) -- (5,0);
    \draw[->] (0,0) -- (0,5);
    \draw (5.5,0) node {$t_0$};
    \draw (0,5.5) node {$t_1$};
    \fill[fill=gray] (1,3) -- (2,3) -- (2,4) -- (1,4) -- cycle;
    \fill[fill=gray] (3,1) -- (4,1) -- (4,2) -- (3,2) -- cycle;
    \draw (1.5,3.5) node {$0$};
    \draw (3.5,1.5) node {$1$};
    \foreach \x in {1,2,3,4} \draw (\x,-1mm) -- (\x,1mm);
    \foreach \x in {1,2,3,4} \draw (-1mm,\x) -- (1mm,\x);
    \draw (1,-.6) node {$x^0_0$};
    \draw (2,-.6) node {$y^0_0$};
    \draw (3,-.6) node {$x^1_0$};
    \draw (4,-.6) node {$y^1_0$};
    \draw (-.6,1) node {$x^1_1$};
    \draw (-.6,2) node {$y^1_1$};
    \draw (-.6,3) node {$x^0_1$};
    \draw (-.6,4) node {$y^0_1$};
  \end{tikzpicture}
  \]
\end{example}

\subsection{The index poset}
\label{indexposet}
Let us come back to the second program of Example~\ref{ex:geom-rel}.  We will
determine the different traces, and their relationships in the trace space, by
combinatorially looking at the way they can turn around holes. To see this in
that example, we extend each hole in parallel to the axes, below or leftwards
from the holes, until they reach the boundary of the state space. These new
obstructions impose traces to go the other way around each hole: the existence
of deadlocks, given these new constraints in the trace space allows us to
determine whether traces going one way or the other around each hole exist. In
fact, this combinatorial information precisely computes all of the trace
space~\cite{raussen2010simplicial}.

In the second program of Example~\ref{ex:geom-rel}, there are four possibilities
to extend once each of the two holes:
\begin{equation}
  \label{eq:ex-hole-ext}
  \begin{tikzpicture}[scale=.3]
    \draw[line width=0.3mm,color=red] (0,0) .. controls (.5,5) .. (5,5);
    \fill[fill=lightgray] (1,0) -- (1,4) -- (2,4) -- (2,0) -- cycle;
    \fill[fill=lightgray] (3,0) -- (3,2) -- (4,2) -- (4,0) -- cycle;
    \fill[fill=gray] (1,3) -- (2,3) -- (2,4) -- (1,4) -- cycle;
    \fill[fill=gray] (3,1) -- (4,1) -- (4,2) -- (3,2) -- cycle;
    \draw[->] (0,0) -- (5,0);
    \draw[->] (0,0) -- (0,5);
    \draw (5.5,0) node {$t_0$};
    \draw (0,5.5) node {$t_1$};
  \end{tikzpicture}
  \begin{tikzpicture}[scale=.3]
    \draw[line width=0.3mm,color=red] (0,0) .. controls (1,3) and (4,3) .. (5,5);
    \fill[fill=lightgray] (0,3) -- (0,4) -- (2,4) -- (2,3) -- cycle;
    \fill[fill=lightgray] (3,0) -- (3,2) -- (4,2) -- (4,0) -- cycle;
    \fill[fill=gray] (1,3) -- (2,3) -- (2,4) -- (1,4) -- cycle;
    \fill[fill=gray] (3,1) -- (4,1) -- (4,2) -- (3,2) -- cycle;
    \draw[->] (0,0) -- (5,0);
    \draw[->] (0,0) -- (0,5);
    \draw (5.5,0) node {$t_0$};
    \draw (0,5.5) node {$t_1$};
  \end{tikzpicture}
  \begin{tikzpicture}[scale=.3]
    \draw[line width=0.3mm,color=red] (0,0) .. controls (5,.5) .. (5,5);
    \fill[fill=lightgray] (0,1) -- (4,1) -- (4,2) -- (0,2) -- cycle;
    \fill[fill=lightgray] (0,3) -- (2,3) -- (2,4) -- (0,4) -- cycle;
    \fill[fill=gray] (1,3) -- (2,3) -- (2,4) -- (1,4) -- cycle;
    \fill[fill=gray] (3,1) -- (4,1) -- (4,2) -- (3,2) -- cycle;
    \draw[->] (0,0) -- (5,0);
    \draw[->] (0,0) -- (0,5);
    \draw (5.5,0) node {$t_0$};
    \draw (0,5.5) node {$t_1$};
  \end{tikzpicture}
  \begin{tikzpicture}[scale=.3]
    \fill[fill=lightgray] (1,0) -- (1,4) -- (2,4) -- (2,0) -- cycle;
    \fill[fill=lightgray,color=lightgray] (0,1) -- (0,2) -- (4,2) -- (4,1) -- cycle;
    \fill[fill=gray] (1,3) -- (2,3) -- (2,4) -- (1,4) -- cycle;
    \fill[fill=gray] (3,1) -- (4,1) -- (4,2) -- (3,2) -- cycle;
    \draw[->] (0,0) -- (5,0);
    \draw[->] (0,0) -- (0,5);
    \draw (5.5,0) node {$t_0$};
    \draw (0,5.5) node {$t_1$};
  \end{tikzpicture}
\end{equation}
Notice that there exists a total path in the first three spaces (as depicted
above), whereas there is none in the last one.

A simple way to encode the combinatorial information about the extension of
holes is through boolean matrices. We write~$\Mat_{l,n}$ for the poset of
$l\times n$ matrices, with~$l$ rows (the number of holes $R^i$) and~$n$ columns
(the dimension of the space, \ie the number of threads in the program), with
coefficients in~$\Z/2\Z$, with the pointwise ordering such that $0\leq 1$: we
have $M\leq N$ whenever
\begin{equation}
  \label{eq:mat-order}
  \forall (i,j)\in\Nint 0l\times\Nint 0n,\qquad M(i,j)\leq N(i,j)
\end{equation}
where~$\Nint mn$ denotes the set $\set{m,\ldots, n-1}$ of integers and $M(i,j)$
denotes the $(i,j)$\nbd{}th coefficient of~$M$. We also write $\Mat_{l,n}^R$ for
the subposet of~$\Mat_{l,n}$ consisting of matrices whose row vectors are all
different from the zero vector, and~$\Mat_{l,n}^C$ for the subposet
of~$\Mat_{l,n}$ consisting of matrices whose column vectors are all unit vectors
(containing exactly one coefficient~$1$).

Given a matrix $M\in\Mat_{l,n}$, we define~$X_M$ as the subspace of~$X$ obtained
by extending downwards each forbidden rectangle~$R^i$ in every direction~$j'$
different from~$j$ for every~$j$ such that~$M(i,j)=1$. Formally,
\[
X_M
\qeq
\dui^n\setminus\bigcup_{M(i,j)=1}\tilde{R}^i_j
\]
where $\tilde{R}^i_j= \prod_{j'=0}^{j-1}[0,y^i_{j'}[\times]x^i_j,y^i_j[
\times\prod_{j'=j+1}^{n-1}[0,y^i_{j'}[$, see~\eqref{eq:ex-hole-ext} and
Example~\ref{ex:mat2d} below.

In order to study whether there is a total path in the space associated to a
matrix, we define a map \hbox{$\Psi:\Mat_{l,n}\to\Z/2\Z$} by~$\Psi(M)=1$ iff
$\tspace{X_M}=\emptyset$, \ie there is no total path in~$X_M$. A matrix~$M$ is
\emph{dead} when $\Psi(M)=1$ and \emph{alive} otherwise. The map~$\Psi$ can
easily be shown to be order preserving.

\begin{definition}
  We write
  \[
  \D(X)
  \qeq
  \setof{M\in\Mat_{l,n}^C}{\Psi(M)=1}
  \]
  for the set of (column) dead matrices and
  \[
  \C(X)
  \qeq
  \setof{M\in\Mat_{l,n}^R}{\Psi(M)=0}
  \]
  for the set of alive matrices (with non-empty rows), which is called the
  \emph{index poset} -- it is implicitly ordered by the
  relation~\eqref{eq:mat-order}.
\end{definition}

\begin{example}
  \label{ex:mat2d}
  In the example above, the three extensions of holes~\eqref{eq:ex-hole-ext} are
  respectively encoded by the following matrices:
  \[
  \left(
    \begin{matrix}
      1&0\\
      1&0\\
    \end{matrix}
  \right)
  \qquad
  \left(
    \begin{matrix}
      0&1\\
      1&0\\
    \end{matrix}
  \right)
  \qquad
  \left(
    \begin{matrix}
      0&1\\
      0&1\\
    \end{matrix}
  \right)
  \qquad
  \left(
    \begin{matrix}
      1&0\\
      0&1\\
    \end{matrix}
  \right)
  \]
  The last matrix is dead and the three others are alive. The last matrix being
  dead indicates that there is no way a trace can pass left of the upper left
  hole and carry on passing below the lower right hole.
\end{example}

\comment{
\begin{example}
  \label{ex:mat3d}
  The geometric semantics of the program constituted of three copies of the
  thread $\P a.\V a.\P b.\V b$ in parallel, with $\capacity a=\capacity b=2$, is
  \vspace{-2ex}
  \[
\begin{tikzpicture}[line join=round,opacity=0.7]
\draw[arrows=->](.786,-.786)--(.435,-.115);
\filldraw[fill=gray](1.479,.73)--(1.794,.743)--(1.819,1.061)--(1.5,1.05)--cycle;
\filldraw[fill=gray](1.908,.652)--(1.935,.982)--(1.819,1.061)--(1.794,.743)--cycle;
\filldraw[fill=gray](1.583,.639)--(1.908,.652)--(1.794,.743)--(1.479,.73)--cycle;
\filldraw[fill=gray](1.583,.639)--(1.605,.97)--(1.5,1.05)--(1.479,.73)--cycle;
\filldraw[fill=gray](1.605,.97)--(1.935,.982)--(1.819,1.061)--(1.5,1.05)--cycle;
\filldraw[fill=gray](1.583,.639)--(1.908,.652)--(1.935,.982)--(1.605,.97)--cycle;
\filldraw[fill=gray](1.068,-.283)--(1.416,-.263)--(1.314,-.133)--(.978,-.151)--cycle;
\filldraw[fill=gray](1.416,-.263)--(1.437,.072)--(1.333,.192)--(1.314,-.133)--cycle;
\filldraw[fill=gray](.978,-.151)--(1.314,-.133)--(1.333,.192)--(.993,.175)--cycle;
\filldraw[fill=gray](1.068,-.283)--(1.084,.055)--(.993,.175)--(.978,-.151)--cycle;
\filldraw[fill=gray](1.084,.055)--(1.437,.072)--(1.333,.192)--(.993,.175)--cycle;
\filldraw[fill=gray](1.068,-.283)--(1.416,-.263)--(1.437,.072)--(1.084,.055)--cycle;
\draw[arrows=->](.786,-.786)--(2.547,-.674);
\draw[arrows=->](.786,-.786)--(.853,1.028);
\draw (2.786,-.659) node {$t_0$};\draw (.86,1.226) node {$t_1$};\draw (.326,.094) node {$t_2$};\draw (1.202,-.042) node {$0$};\draw (1.702,.853) node {$1$};\end{tikzpicture}
  \]
  The spaces~$X_M$ corresponding to the matrices
  \[
  \left(
    \begin{matrix}
      1&0&0\\
      0&0&1\\
    \end{matrix}
  \right)
  \qquad
  \left(
    \begin{matrix}
      0&0&1\\
      1&0&0\\
    \end{matrix}
  \right)
  \qquad
  \left(
    \begin{matrix}
      0&0&0\\
      1&1&1\\
    \end{matrix}
  \right)
  \]
  are respectively
  \[
\begin{tikzpicture}[line join=round,opacity=0.7]
\draw[arrows=->](.786,-.786)--(.435,-.115);
\filldraw[fill=gray](1.479,.73)--(1.794,.743)--(1.819,1.061)--(1.5,1.05)--cycle;
\filldraw[fill=lightgray](.496,-.231)--(1.723,-.164)--(1.819,1.061)--(.524,1.017)--cycle;
\filldraw[fill=lightgray](1.831,-.283)--(1.935,.982)--(1.819,1.061)--(1.723,-.164)--cycle;
\filldraw[fill=lightgray](.561,-.354)--(1.831,-.283)--(1.723,-.164)--(.496,-.231)--cycle;
\filldraw[fill=lightgray](.561,-.354)--(.593,.934)--(.524,1.017)--(.496,-.231)--cycle;
\draw[line width=0.5mm,color=red](1.999,1.148)--(2.042,1.461);
\filldraw[fill=gray](1.908,.652)--(1.935,.982)--(1.819,1.061)--(1.794,.743)--cycle;
\filldraw[fill=gray](1.583,.639)--(1.908,.652)--(1.794,.743)--(1.479,.73)--cycle;
\filldraw[fill=gray](1.583,.639)--(1.605,.97)--(1.5,1.05)--(1.479,.73)--cycle;
\filldraw[fill=lightgray](.561,-.354)--(1.831,-.283)--(1.935,.982)--(.593,.934)--cycle;
\filldraw[fill=gray](1.605,.97)--(1.935,.982)--(1.819,1.061)--(1.5,1.05)--cycle;
\filldraw[fill=lightgray](.593,.934)--(1.935,.982)--(1.819,1.061)--(.524,1.017)--cycle;
\draw[line width=0.5mm,color=red](1.954,.817)--(1.999,1.148);
\filldraw[fill=gray](1.583,.639)--(1.908,.652)--(1.935,.982)--(1.605,.97)--cycle;
\draw[line width=0.5mm,color=red](.936,-.089)--(1.954,.817);
\filldraw[fill=lightgray](.964,-.468)--(1.295,-.448)--(1.333,.192)--(.993,.175)--cycle;
\filldraw[fill=gray](1.416,-.263)--(1.437,.072)--(1.333,.192)--(1.314,-.133)--cycle;
\filldraw[fill=lightgray](1.503,-.741)--(1.55,-.057)--(1.333,.192)--(1.295,-.448)--cycle;
\filldraw[fill=lightgray](1.147,-.763)--(1.503,-.741)--(1.295,-.448)--(.964,-.468)--cycle;
\draw[line width=0.5mm,color=red](.908,-.219)--(.936,-.089);
\filldraw[fill=gray](1.068,-.283)--(1.416,-.263)--(1.314,-.133)--(.978,-.151)--cycle;
\filldraw[fill=gray](.978,-.151)--(1.314,-.133)--(1.333,.192)--(.993,.175)--cycle;
\filldraw[fill=gray](1.068,-.283)--(1.416,-.263)--(1.437,.072)--(1.084,.055)--cycle;
\filldraw[fill=lightgray](1.147,-.763)--(1.183,-.076)--(.993,.175)--(.964,-.468)--cycle;
\draw[line width=0.5mm,color=red](.786,-.786)--(.908,-.219);
\filldraw[fill=gray](1.068,-.283)--(1.084,.055)--(.993,.175)--(.978,-.151)--cycle;
\filldraw[fill=gray](1.084,.055)--(1.437,.072)--(1.333,.192)--(.993,.175)--cycle;
\filldraw[fill=lightgray](1.183,-.076)--(1.55,-.057)--(1.333,.192)--(.993,.175)--cycle;
\draw[arrows=->](.786,-.786)--(2.547,-.674);
\filldraw[fill=lightgray](1.147,-.763)--(1.503,-.741)--(1.55,-.057)--(1.183,-.076)--cycle;
\draw[arrows=->](.786,-.786)--(.853,1.028);
\draw (2.786,-.659) node {$t_0$};\draw (.86,1.226) node {$t_1$};\draw (.326,.094) node {$t_2$};\end{tikzpicture}
  \hspace{-1ex}
\begin{tikzpicture}[line join=round,opacity=0.7]
\draw[arrows=->](.786,-.786)--(.435,-.115);
\draw[line width=0.5mm,color=red](1.402,.316)--(2.042,1.461);
\filldraw[fill=lightgray](1.421,-.18)--(1.723,-.164)--(1.819,1.061)--(1.5,1.05)--cycle;
\filldraw[fill=lightgray](2.203,-.696)--(2.345,.703)--(1.819,1.061)--(1.723,-.164)--cycle;
\filldraw[fill=lightgray](1.855,-.718)--(2.203,-.696)--(1.723,-.164)--(1.421,-.18)--cycle;
\filldraw[fill=gray](1.908,.652)--(1.935,.982)--(1.819,1.061)--(1.794,.743)--cycle;
\filldraw[fill=gray](1.583,.639)--(1.908,.652)--(1.794,.743)--(1.479,.73)--cycle;
\filldraw[fill=gray](1.479,.73)--(1.794,.743)--(1.819,1.061)--(1.5,1.05)--cycle;
\filldraw[fill=gray](1.583,.639)--(1.908,.652)--(1.935,.982)--(1.605,.97)--cycle;
\filldraw[fill=lightgray](1.855,-.718)--(1.976,.687)--(1.5,1.05)--(1.421,-.18)--cycle;
\draw[line width=0.5mm,color=red](1.41,.245)--(1.402,.316);
\draw[line width=0.5mm,color=red](1.46,-.24)--(1.41,.245);
\filldraw[fill=gray](1.583,.639)--(1.605,.97)--(1.5,1.05)--(1.479,.73)--cycle;
\filldraw[fill=gray](1.605,.97)--(1.935,.982)--(1.819,1.061)--(1.5,1.05)--cycle;
\filldraw[fill=lightgray](1.976,.687)--(2.345,.703)--(1.819,1.061)--(1.5,1.05)--cycle;
\filldraw[fill=gray](1.416,-.263)--(1.437,.072)--(1.333,.192)--(1.314,-.133)--cycle;
\filldraw[fill=lightgray](1.395,-.589)--(1.437,.072)--(1.333,.192)--(1.295,-.448)--cycle;
\filldraw[fill=lightgray](.63,-.487)--(1.295,-.448)--(1.333,.192)--(.649,.158)--cycle;
\filldraw[fill=lightgray](.705,-.631)--(1.395,-.589)--(1.295,-.448)--(.63,-.487)--cycle;
\filldraw[fill=lightgray](.705,-.631)--(.727,.037)--(.649,.158)--(.63,-.487)--cycle;
\filldraw[fill=gray](1.068,-.283)--(1.416,-.263)--(1.314,-.133)--(.978,-.151)--cycle;
\filldraw[fill=gray](.978,-.151)--(1.314,-.133)--(1.333,.192)--(.993,.175)--cycle;
\filldraw[fill=gray](1.068,-.283)--(1.084,.055)--(.993,.175)--(.978,-.151)--cycle;
\filldraw[fill=gray](1.084,.055)--(1.437,.072)--(1.333,.192)--(.993,.175)--cycle;
\filldraw[fill=lightgray](.727,.037)--(1.437,.072)--(1.333,.192)--(.649,.158)--cycle;
\filldraw[fill=lightgray](.705,-.631)--(1.395,-.589)--(1.437,.072)--(.727,.037)--cycle;
\filldraw[fill=gray](1.068,-.283)--(1.416,-.263)--(1.437,.072)--(1.084,.055)--cycle;
\draw[line width=0.5mm,color=red](1.47,-.332)--(1.46,-.24);
\draw[arrows=->](.786,-.786)--(2.547,-.674);
\filldraw[fill=lightgray](1.855,-.718)--(2.203,-.696)--(2.345,.703)--(1.976,.687)--cycle;
\draw[line width=0.5mm,color=red](.786,-.786)--(1.47,-.332);
\draw[arrows=->](.786,-.786)--(.853,1.028);
\draw (2.786,-.659) node {$t_0$};\draw (.86,1.226) node {$t_1$};\draw (.326,.094) node {$t_2$};\end{tikzpicture}
  \hspace{-1ex}
\begin{tikzpicture}[line join=round,opacity=0.7]
\draw[arrows=->](.786,-.786)--(.435,-.115);
\filldraw[fill=lightgray](1.421,-.18)--(1.723,-.164)--(1.819,1.061)--(1.5,1.05)--cycle;
\filldraw[fill=lightgray](2.203,-.696)--(2.345,.703)--(1.819,1.061)--(1.723,-.164)--cycle;
\filldraw[fill=lightgray](1.855,-.718)--(2.203,-.696)--(1.723,-.164)--(1.421,-.18)--cycle;
\filldraw[fill=gray](1.479,.73)--(1.794,.743)--(1.819,1.061)--(1.5,1.05)--cycle;
\filldraw[fill=lightgray](.496,-.231)--(1.723,-.164)--(1.819,1.061)--(.524,1.017)--cycle;
\filldraw[fill=lightgray](1.831,-.283)--(1.935,.982)--(1.819,1.061)--(1.723,-.164)--cycle;
\filldraw[fill=lightgray](.561,-.354)--(1.831,-.283)--(1.723,-.164)--(.496,-.231)--cycle;
\filldraw[fill=lightgray,draw=none](1.583,.639)--(1.518,-.3)--(1.831,-.283)--(1.908,.652)--cycle;
\draw(1.518,-.3)--(1.831,-.283)--(1.908,.652);
\filldraw[fill=gray](1.583,.639)--(1.908,.652)--(1.794,.743)--(1.479,.73)--cycle;
\filldraw[fill=lightgray,draw=none](1.583,.639)--(1.908,.652)--(1.794,.743)--(1.479,.73)--cycle;
\draw(1.908,.652)--(1.794,.743)--(1.479,.73);
\filldraw[fill=lightgray](2.308,.336)--(2.345,.703)--(1.819,1.061)--(1.794,.743)--cycle;
\filldraw[fill=lightgray](.517,.692)--(1.794,.743)--(1.819,1.061)--(.524,1.017)--cycle;
\filldraw[fill=gray](1.908,.652)--(1.935,.982)--(1.819,1.061)--(1.794,.743)--cycle;
\filldraw[fill=lightgray,draw=none](1.583,.639)--(1.908,.652)--(1.935,.982)--(1.605,.97)--cycle;
\draw(1.908,.652)--(1.935,.982)--(1.605,.97);
\filldraw[fill=gray](1.583,.639)--(1.908,.652)--(1.935,.982)--(1.605,.97)--cycle;
\filldraw[fill=lightgray,draw=none](1.583,.639)--(1.944,.319)--(2.308,.336)--(1.908,.652)--cycle;
\draw(1.944,.319)--(2.308,.336)--(1.908,.652);
\filldraw[fill=lightgray](1.855,-.718)--(1.976,.687)--(1.5,1.05)--(1.421,-.18)--cycle;
\filldraw[fill=lightgray,draw=none](.585,.598)--(1.583,.639)--(1.479,.73)--(.517,.692)--cycle;
\draw(1.479,.73)--(.517,.692)--(.585,.598);
\filldraw[fill=lightgray](.561,-.354)--(.593,.934)--(.524,1.017)--(.496,-.231)--cycle;
\filldraw[fill=gray](1.583,.639)--(1.605,.97)--(1.5,1.05)--(1.479,.73)--cycle;
\filldraw[fill=lightgray,draw=none](.561,-.354)--(1.518,-.3)--(1.605,.97)--(.593,.934)--cycle;
\draw(1.605,.97)--(.593,.934)--(.561,-.354)--(1.518,-.3);
\filldraw[fill=lightgray](.839,.641)--(2.345,.703)--(1.819,1.061)--(.524,1.017)--cycle;
\filldraw[fill=gray](1.605,.97)--(1.935,.982)--(1.819,1.061)--(1.5,1.05)--cycle;
\filldraw[fill=lightgray](1.976,.687)--(2.345,.703)--(1.819,1.061)--(1.5,1.05)--cycle;
\filldraw[fill=lightgray](.593,.934)--(1.935,.982)--(1.819,1.061)--(.524,1.017)--cycle;
\filldraw[fill=lightgray,draw=none](.585,.598)--(.825,.267)--(1.944,.319)--(1.583,.639)--cycle;
\draw(.585,.598)--(.825,.267)--(1.944,.319);
\filldraw[fill=lightgray](.825,.267)--(.839,.641)--(.524,1.017)--(.517,.692)--cycle;
\filldraw[fill=gray](1.068,-.283)--(1.416,-.263)--(1.314,-.133)--(.978,-.151)--cycle;
\filldraw[fill=gray](1.416,-.263)--(1.437,.072)--(1.333,.192)--(1.314,-.133)--cycle;
\filldraw[fill=gray](.978,-.151)--(1.314,-.133)--(1.333,.192)--(.993,.175)--cycle;
\filldraw[fill=gray](1.068,-.283)--(1.084,.055)--(.993,.175)--(.978,-.151)--cycle;
\filldraw[fill=gray](1.084,.055)--(1.437,.072)--(1.333,.192)--(.993,.175)--cycle;
\filldraw[fill=gray](1.068,-.283)--(1.416,-.263)--(1.437,.072)--(1.084,.055)--cycle;
\draw[arrows=->](.786,-.786)--(2.547,-.674);
\filldraw[fill=lightgray](1.855,-.718)--(2.203,-.696)--(2.345,.703)--(1.976,.687)--cycle;
\draw[arrows=-](.786,-.786)--(.825,.267);
\filldraw[fill=lightgray](.825,.267)--(2.308,.336)--(2.345,.703)--(.839,.641)--cycle;
\draw[arrows=-](.825,.267)--(.839,.641);
\draw[arrows=->](.839,.641)--(.853,1.028);
\draw (2.786,-.659) node {$t_0$};\draw (.86,1.226) node {$t_1$};\draw (.326,.094) node {$t_2$};\end{tikzpicture}
  \]
  The first two matrices are alive, as shown by the drawn total paths.
\end{example}
}

A reason why the matrices in the index poset are convenient objects to study the
schedulings is that they are topologically very
simple~\cite{raussen2010simplicial}:
\begin{proposition}
  \label{prop:alive-dihom}
  For any matrix $M\in\Mat_{l,n}^R$, the space $X_M(x,y)$ is either empty or
  contractible: any two paths with the same source~$x$ and target~$y$ are
  dihomotopic. In particular, for any matrix $M\in\C(X)$, the space
  $X_M(0,\infty)$ is always contractible.
\end{proposition}

Our main interest in the index poset is that it enables us to compute the
schedulings (\ie maximal paths modulo dihomotopy) of the space: these
schedulings are in bijection with alive matrices in~$\C(X)$ modulo an
equivalence relation called \emph{connexity}, which is defined as follows. Given
two matrices~$M,N\in\Mat_{l,n}$, their \emph{intersection} $M\wedge N$ is
defined as the matrix~$M\wedge N$ such that $(M\wedge
N)(i,j)=\min(M(i,j),N(i,j))$.

\begin{definition}
  \label{def:connected}
  Two matrices~$M$ and~$N$ are \emph{connected} when their intersection does not
  contain any row filled with~$0$.
\end{definition}

\noindent
The dihomotopy classes of total paths in~$X$ can finally be computed thanks to
the following property:

\begin{proposition}
  \label{prop:conn-comp}
  The connected components of~$\C(X)$ are in bijection with schedulings in~$X$.
\end{proposition}

\begin{example}
  \label{ex:cube}
  Consider the program $p=q|q|q$ where $q=\P a.\V a$. The associated trace
  space~$X_p$ is a cube minus a cube (as shown in Example~\ref{ex:delooping}).
  The matrices in~$\C(X_p)$ are
  \[
  \begin{array}{c@{\qquad}c@{\qquad}c@{\qquad}c@{\qquad}c@{\qquad}c}
    \pa{
      \begin{matrix}
        1&0&0
      \end{matrix}
    }
    &
    \pa{
      \begin{matrix}
        0&1&0
      \end{matrix}
    }
    &
    \pa{
      \begin{matrix}
        0&0&1
      \end{matrix}
    }
    &
    \pa{
      \begin{matrix}
        0&1&1
      \end{matrix}
    }
    &
    \pa{
      \begin{matrix}
        1&0&1
      \end{matrix}
    }
    &
    \pa{
      \begin{matrix}
        1&1&0
      \end{matrix}
    }
  \end{array}
  \]
  and they are all (transitively) connected. For instance,
  $
  \pa{
    \begin{matrix}
      0&1&1
    \end{matrix}
  }
  \wedge
  \pa{
    \begin{matrix}
      1&0&1
    \end{matrix}
  }
  =
  \pa{
    \begin{matrix}
      0&0&1
    \end{matrix}
  } $. The program~$p$ thus has exactly one total scheduling, as expected.
\end{example}

Intuitively, alive matrices describe sets of dihomotopic total paths
(Proposition~\ref{prop:alive-dihom}) and the fact that two matrices have
non-zero rows in their intersection means that there are paths which satisfy
the constraints imposed by both matrices, \ie the two matrices describe the same
dihomotopy class of total paths.

\subsection{Computing dihomotopy classes}
\label{sec:homotopy-classes}
\label{algo}
The computation of the dihomotopy classes of total paths in the geometric
semantics~$X$ of a given program will be performed in three steps:
\begin{enumerate}
\item we compute the set $\D(X)$ of dead matrices,
\item we use~$\D(X)$ to compute the index poset~$\C(X)$,
\item we deduce the homotopy classes of total paths by quotienting~$\C(X)$ by
  the connexity relation.
\end{enumerate}
These steps are detailed below.




Given a subset~$I$ of $\Nint 0l$ and an index~$j\in\Nint 0n$, we
write~$y^I_j=\min\setof{y^i_j}{i\in I}$ (by convention
\hbox{$y^\emptyset_j=\infty$}). Given a matrix~$M\in \Mat_{l,n}$, we define the
set of \emph{non-zero rows} of~$M$ by \hbox{$R(M)=\setof{i\in\Nint 0l}{\exists
    j\in\Nint 0n,\ M(i,j)\neq 0}$}. It can be shown that a matrix~$M$ is dead if
and only if the space $X_M$ contains a deadlock. From the characterization of
deadlocks in geometric semantics given in~\cite{fajstrup1998detecting}, the
following characterization of dead matrices can therefore be deduced:



\begin{proposition}
  A matrix $M\in \Mat_{l,n}^C$ is in $\D(X)$ iff it satisfies
  \begin{equation}
    \label{eq:dead}
    \forall (i,j)\in\Nint 0l\times\Nint 0n,
    \quad
    M(i,j)=1
    \Rightarrow
    x^i_j<y^{R(M)}_j
  \end{equation}
\end{proposition}

\begin{example}
  In the example below with~$l=2$ and~$n=2$, the matrix~$M=\left(
    \begin{matrix}
      0&1\\
      1&0\\
    \end{matrix}
  \right)$ is dead (we suppose that $x^i_j=1+i(j+1)$ and $y^i_j=3+i(j+1)-j$):
  \[
  \vcenter{
    \begin{tikzpicture}[scale=.5]
      \draw (5.5,0) node {$t_0$};
      \draw (-.6,5) node {$t_1$};
      \fill[fill=lightgray] (0,1) -- (0,2) -- (3,2) -- (3,1) -- cycle;
      \fill[fill=lightgray] (2,0) -- (2,4) -- (4,4) -- (4,0) -- cycle;
      \fill[fill=gray] (1,1) -- (3,1) -- (3,2) -- (1,2) -- cycle;
      \fill[fill=gray] (2,3) -- (4,3) -- (4,4) -- (2,4) -- cycle;
      \draw (2,1.5) node {$0$};
      \draw (3,3.5) node {$1$};
      \foreach \x in {1,2,3,4} \draw (\x,-1mm) -- (\x,1mm);
      \foreach \x in {1,2,3,4} \draw (-1mm,\x) -- (1mm,\x);
      \draw (1,-.6) node {$x^0_0$};
      \draw (2,-.6) node {$x^1_0$};
      \draw (3,-.6) node {$y^0_0$};
      \draw (4,-.6) node {$y^1_0$};
      \draw (-.6,1) node {$x^0_1$};
      \draw (-.6,2) node {$y^0_1$};
      \draw (-.6,3) node {$x^1_1$};
      \draw (-.6,4) node {$y^1_1$};
      \draw[->] (0,0) -- (5,0);
      \draw[->] (0,0) -- (0,5);
    \end{tikzpicture}
  }
  \hspace{-30ex}
  \begin{array}{c}
    x^0_1=1<2=y^{\set{0,1}}_1\\
    x^1_0=2<3=y^{\set{0,1}}_0
  \end{array}
  \]
\end{example}

The above proposition enables us to compute the set of dead matrices, for
instance by enumerating all matrices and checking whether they satisfy
condition~\ref{eq:dead} (a more efficient method is described in
Section~\ref{sec:implem}). From this set, the index poset~$\C(X)$ can be
determined using the following property:

\begin{proposition}
  \label{prop:index-dead}
  A matrix $M\in \Mat_{l,n}$ is not in~$\C(X)$ iff there exists a matrix
  \hbox{$N\in \D(X)$} such that $N\leq M$. In other words, $M\in\C(X)$ iff for
  every matrix \hbox{$N\in \D(X)$} there exists indexes $i\in\Nint 0l$ and
  $j\in\Nint 0n$ such that $M(i,j)=0$ and~$N(i,j)=1$.
\end{proposition}

Notice that the poset~$\C(X)$ is downward closed (because~$\Psi$ is order
preserving) and one is naturally interested in the subset~$\C_{\max}(X)$ of
\emph{maximal} matrices in order to describe
it. Proposition~\ref{prop:index-dead} provides a simple-minded algorithm for
computing (maximal) matrices in~$\C(X)$. We write
$\D(X)=\set{D_0,\ldots,D_{p-1}}$. We then compute the sets~$C_k$ of maximal
matrices~$M$ such that for every $i\in\Nint 0k$ we have~$D_i\not\leq M$. We
start from the set \hbox{$C_0=\set{\fatone}$} where~$\fatone$ is the matrix
containing only~$1$ as coefficients. Given a matrix~$M$, we
write~$M^{\lnot(i,j)}$ for the matrix obtained from~$M$ by replacing the
$(i,j)$\nbd{}th coefficient by $1-M(i,j)$. The set~$C_{k+1}$ is then computed
from~$C_k$ by doing the following for all matrices~$M\in C_k$ such that
\hbox{$D_k\leq M$}:
\begin{enumerate}
\item remove~$M$ from~$C_k$,
\item for every~$(i,j)$ such that~$D_k(i,j)=1$,
  \begin{itemize}
  \item remove every matrix~$N\in C_k$ such that~$N\leq
    M^{\lnot(i,j)}$,
  \item if there exists no matrix~$N\in C_k$ such that~$M^{\lnot(i,j)}\leq N$,
    add~$M^{\lnot(i,j)}$ to~$C_k$.
  \end{itemize}
\end{enumerate}
The set~$\C_{\max}(X)$ is obtained as $C_p$. If we remove the second point and
replace it by
\begin{enumerate}
\item[2'.] for every~$(i,j)$ such that~$D_k(i,j)=1$ and
  ~$M^{\lnot(i,j)}\in\Mat^R_{l,n}$, \hbox{add~$M^{\lnot(i,j)}$ to~$C_k$.}
\end{enumerate}
we compute a set~$C_p$ such that~$\C_{\max}(X)\subseteq C_p\subseteq\C(X)$,
which is enough to compute connected components and has proved faster to compute
in practice.

\begin{example}
  \label{exnaive}
  Consider again Example \ref{ex:geom-rel}.
The algorithm starts with
  \[
  \small
  C_0
  \qeq
  \left\{
    M_0=
    \pa{
      \begin{matrix}
        1&1\\
        1&1
      \end{matrix}
    }
  \right\}
  \]
  For $C_1$, we must have $D_0\not\leq M_0$ so we swap any of the two ones in the
  first row:
  \[
  \small
  C_1
  \qeq
  \left\{
    M_1=
    \pa{
      \begin{matrix}
        0 & 1 \\
        1 & 1
      \end{matrix}
    },
    M_2=
    \pa{
      \begin{matrix}
        1 & 0 \\
        1 & 1
      \end{matrix}
    }
  \right\}
  \]
  Similarly for $C_2$, we have to swap the bits on the second row so that
  $D_1\not\leq M_i$:
  \[
  C_2=
  \left\{
    M_3=
    \pa{
      \begin{matrix}
        0 &  1 \\
        0 &  1
      \end{matrix}
    },
    M_4=\pa{
      \begin{matrix}
        0 &  1 \\
        1 &  0
      \end{matrix}
    },
    M_5=\pa{
      \begin{matrix}
        1 &  0 \\
        0 &  1
      \end{matrix}
    },
    M_6=\pa{
      \begin{matrix}
        1 &  0 \\
        1 &  0
      \end{matrix}
    }
  \right\}
  \]
  Finally, we have $D_2\not\leq M_i$, excepting $D_2\leq M_5$, so we swap the
  bits in position $(1,1)$ and in position $(2,2)$:
  \[
  \small
  M_5'=\pa{
    \begin{matrix}
      0 & 0 \\
      0 & 1
      \end{matrix}
    }
    \leq M_3
    \qquad\qquad
    M_5''=\pa{
      \begin{matrix}
        1 & 0 \\
        0 & 0
      \end{matrix}
    }
    \leq M_6
  \]
  Since we are only interested in maximal matrices, we end up with
  $C_3=\set{M_6,M_4,M_3}$. The trace spaces corresponding to those matrices are
  the three first depicted in~\eqref{eq:ex-hole-ext}. None of those matrices
  being connected, the trace space up to dihomotopy consists of exactly 3
  distinct points.
\end{example}

Other implementations of the algorithm can be obtained by reformulating the
computation of~$\C_{\max}(X)$ as finding a minimal transversal in a
hypergraph, 
for which efficient algorithms have been
proposed~\cite{kavvadias1999evaluation}.

We have supposed up to now that the forbidden region was a
union of rectangles~$R^i$, each such rectangle being a product of open
intervals~$\dui^i_j=]x^i_j,y^i_j[$. The algorithm given above can easily be
generalized to the case where the rectangles~$R^i$ can ``touch the boundary'' in
some dimensions, \ie the intervals~$\dui^i_j$ are either of the form
$]x^i_j,y^i_j[$ or $[0,y^i_j[$ or $]x^i_j,\infty]$ or $[0,\infty]$. For example,
the process $\P a.\V a|\P a.\V a|\P a.\V a$, with~$\capacity a=1$, generates
such a forbidden region. We write~$B\in \Mat_{l,n}$ for the \emph{boundary
  matrix}, which is the matrix such that $B(i,j)=0$ whenever $x^i_j=0$ (\ie the
$i$-th interval touches the lowest boundary in dimension~$j$) and
\hbox{$B(i,j)=1$} otherwise. The matrices of~$\D(X)$ are the matrices~$M\in
\Mat_{n,l}$ of the form \hbox{$M=N\wedge B$}, for some matrix~$N\in M^C_{n,l}$,
which satisfy~\eqref{eq:dead} and such that
\begin{equation}
  \label{eq:dead-bounds}
  \forall j\in C(M),
  \qquad
  y^{R(M)}_j=\infty
\end{equation}
where~$C(M)$ is the set of indexes of null columns of~$M$.

\subsection{An efficient implementation}
\label{sec:implem}
In order to compute the set~$\D(X)$ of dead matrices, the general idea is to
enumerate all the matrices~$M\in\Mat^C_{l,n}$ and check whether they satisfy the
condition~\eqref{eq:dead}. Of course, a direct implementation of this idea would
be highly inefficient since there are $l^n$ matrices in~$\Mat^C_{l,n}$. In order
to improve this, we try to detect ``as soon as possible'' when a matrix does not
satisfy the condition: we first fix the coefficient in the first column of~$M$
and check whether it is possible for a matrix with this first column to be dead,
then we fix the second column and so on. In fact, we have to check that every
coefficient $(i,j)$ such that~$M(i,j)=1$ satisfies~$x^i_j<y^{R(M)}_j$. Now,
suppose that we know some of the coefficients $(i,j)$ for which~$M(i,j)=1$. We
therefore know a subset~$I\subseteq R(M)$ of the non-zero rows. If for one of
these coefficients we have~$x^i_j\geq y^I_j$, we know that the matrix cannot
satisfy the condition~\eqref{eq:dead} because~$x^i_j\geq y^I_j\geq
y^{R(M)}_j$. A similar reasoning can be held for
condition~\eqref{eq:dead-bounds}.

\begin{figure}[t]
{  \centering
{\small
\begin{lstlisting}
let rec compute_dead $j$ $m$ $rows$ $yrows$ =
  if $j$ = $n$ then $dead$ := $m$ :: !$dead$ else
    for $i$ = $0$ to $l$ - 1 do
      try
        let $changed\_rows$ = not (Set.mem $i$ $rows$) in
        let $rows$ = Set.add $i$ $rows$ in
        let $m$ = Array.copy $m$ in
        if $bounds$($i$,$j$) = 1 then $m$.($j$) <- None else $m$.($j$) <- Some $i$;
        (match $m$.($j$) with
            | Some $i$ -> if xij >= $yrows$.($j$) then raise Exit
            | None -> if $yrows$.($j$) <> infty $\ $then raise Exit);
        let $yrows$ =
          let $j'$ = $j$ in
          if not $changed\_rows$ then $yrows$ else
            Array.mapi (fun $j$ yrj ->
                if yrj <= yij then yrj else
                  match $m$.($j$) with
                    | None ->
                        if $j$ <= $j'$ && yij <> infty then raise Exit; yij
                    | Some $i$ ->
                        if xij >= yij then raise Exit; yij
                        ) $yrows$
        in
        compute_dead ($j$+1) $m$ $rows$ $yrows$
      with Exit -> ()
    done
\end{lstlisting}
}
}
 \vspace{-3ex}
 \caption{Algorithm for computing dead matrices.}
 \label{fig:dead-algo}
 \vspace{-4ex}
\end{figure}

The actual function computing the dead matrices is presented in
Figure~\ref{fig:dead-algo}, in pseudo-OCaml code. This recursive function
fills~$j$-th column of the matrix~$M$ (whose columns with index below~$j$ are
supposed to be already fixed) and performs the check: it tries to set the $i$-th
coefficient to~$1$ (and all the others to~$0$) for every~$i\in\Nint 0l$. If a
matrix beginning as~$M$ (up to the $j$-th column) cannot be dead, the
computation is aborted by raising the Exit exception. When all the columns have
been computed the matrix is added to the list $dead$ of dead matrices. Since a
matrix~$M\in\Mat^C_{l,n}$ has at most one non-null coefficient in a given
column, it will be coded as an array of length~$n$ whose $j$-th element is
either None when all the elements of the $j$-th column are null, or Some~$i$
when the $i$-th coefficient of the $j$-th column is~$1$ and the others are
$0$. The argument $rows$ is the set of indexes of known non-null rows of~$M$ and
$yrows$ is an array of length~$n$ such that $yrows$.($j$)$=y^{rows}_j$. The
matrix $bounds$ is the matrix previously noted~$B$ used to perform the
check~\eqref{eq:dead-bounds}.
Notice that the algorithm takes advantage of the
fact that when the coefficient~$i$ chosen for the $j$-th column is already
in~$rows$ (\ie when the variable $changed\_rows$ is false) then many
computations can be spared because the coefficients~$y^{rows}_j$ are not
changed.


Once the set of dead matrices computed, the set~$\C(X)$ of alive matrices is
then computed using the naive algorithm of Section~\ref{sec:homotopy-classes},
exemplified in Example \ref{exnaive}. We have also implemented a simple
hypergraph transversal algorithm \cite{Berge} but it did not bring significant
improvements, more elaborate algorithms might give better results
though. Finally, the representatives of traces are computed as the connected
components (in the sense of Proposition~\ref{prop:conn-comp}) of~$\C(X)$, in a
straightforward way. An explicit sequence of instructions corresponding to every
representative~$M$ can easily be computed: it corresponds to the sequence of
instructions crossed by any increasing total path in the d-space~$X_M$.

\subsection{An example: the $n$ dining philosophers}
\label{benchmarks}
In order to illustrate the performances of our algorithm, we present below the
computation times for the well-known $n$~dining philosophers
program~\cite{philosophers} whose schedulings are 
in $O(2^n)$, hence is pushing any algorithm that would determine the essential
schedules to its (exponential) limits.
It is constituted of~$n$
processes~$p_k$ in parallel, using~$n$ mutexes~$a_i$, defined by
\hbox{$p_k=\P{a_{k}}.\P{a_{k+1}}.\V{a_{k}}.\V{a_{k+1}}$}, where the indexes on
mutexes~$a_i$ are taken modulo~$n$. Such a program generates~$2^n-2$ distinct
schedulings, which our program finds correctly. The table below summarizes the
execution time and memory consumption for our tool \hbox{ALCOOL} (programmed in
OCaml), as well as for the model checker SPIN~\cite{spin} implementing partial
order reduction techniques. Whereas SPIN is not significantly slower, it consumes
much more memory and starts to use swap from $n=12$ (thus failing to give an
answer in a reasonable time for $n>12$). Notice that the implementation of SPIN
is finely tuned and also benefits from \texttt{gcc} optimizations, whereas there
is room for many improvements in ALCOOL. In particular, most of the time is
spent in computing dead matrices and the algorithm of Section~\ref{sec:implem}
could be improved by finding a heuristic to suitably sort holes so that failures
to satisfy condition~\eqref{eq:dead} are detected earlier. The present algorithm
is also significantly faster than some of the author's previous
contribution~\cite{concur05}: for instance, it was unable to generate these
maximal dipaths because of memory requirements, for $n$ philosophers with $n>8$
(in the benchmarks of~\cite{concur05}, it was taking already 13739s, on a 1GHz
laptop computer though, to generate just the component category for 9
philosophers).

\begin{center}
  \begin{tabular}{r|r|r|r|r|r}
    $n$&sched.& ALC. (s)&ALC. (MB)&SP. (s)&SP. (MB)\\
    \hline
    10&1022&5&4&8&179\\
    11&2046&32&9&42&816\\
    12&4094&227&26&313&3508\\
    13&8190&1681&58&$\infty$&$\infty$\\
    14&16382&13105&143&$\infty$&$\infty$\\
  \end{tabular}
\end{center}


Since the size of the output is generally exponential in the size of the input,
there is no hope to find an algorithm which has less than an exponential
worst-case complexity (which our algorithm clearly has). However, since our goal
is to program actual tools to very concurrent programs, practical improvements
in the execution time or memory consumption are really interesting from this
point of view. We have of course tried our tool on many more examples, which
confirm the improvement trend, and shall be presented in a longer version of the
article.



\section{Programs with loops}
\label{programswithloops}
\subsection{Paths in deloopings}
One of the most challenging part of verifying concurrent programs consists in
verifying programs with loops since those contain a priori an infinite number of
possible execution traces. We extend here the previous methodology and, given a
program containing loops, we compute a (finite!) automaton whose accepted paths
describe the schedulings of the program: this automaton, can thus be considered
as a control flow graph of the concurrent program. Of course, we are then able
to use the traditional methods in static analysis, such as abstract
interpretation, to study the program (this is briefly presented in
Section~\ref{sec:static-anal}). This section builds on some ideas being
currently developed by Fajstrup~\cite{LF2011}, however most of the properties
presented in this section are entirely new. To the best of our knowledge, this
is the first works in which geometric methods are used in order devise a
practical algorithm to handle programs containing loops. A particularly
interesting feature of our method lies in the fact that it consider the broad
``geometry of holes'' and can thus associate a small control flow graph to a
given program, see Section~\ref{sec:loops-implem}.

In the following, we suppose fixed a program of the form
$p=p_0|p_1|\ldots|p_{n-1}$ as in~\eqref{eq:prog}, with $n$ threads. We write
\[
p^*\qeq p_0^*\ |\ p_1^*\ |\ \ldots\ |\ p_{n-1}^*
\]
for the associated ``looping program''. Our goal in this section is to describe
the schedulings of such a program~$p^*$ (the restriction on the form of the
programs considered here was only done to simplify our presentation and the
methodology can be extended to handle all well-bracketed programs generated by
the grammar, without any essential technical difficulty added). Following
Section~\ref{geomsem}, its geometrical semantics consists of an
$n$\nbd{}dimensional torus with rectangular holes. As previously, for
simplicity, we suppose that these holes do not intersect the boundaries, \ie
that $p$ satisfies the hypothesis of Section~\ref{tracespaces}. Given an
$n$-dimensional vector~$v=(v_0,\ldots,v_{n-1})$ with coefficients in~$\N$, the
\emph{$v$-delooping} of~$p$, written~$p^v$, is the program
$p_0^{v_0}|p_1^{v_1}|\ldots|p_{n-1}^{v_{n-1}}$, where~$p_j^{v_j}$ denotes the
concatenation of~$v_j$ copies of~$p_j$. A \emph{scheduling} in~$p$ is a
scheduling in the previous sense (\ie a total path modulo homotopy) in~$p^v$ for
some vector~$v$.

\begin{example}
  \label{ex:delooping}
  Consider the program $p=q|q|q$ of Example~\ref{ex:cube}, where $q=\P a.\V
  a$. Its geometric realization~$X_p$ is pictured on the left, and its
  $(3,2,2)$\nbd{}delooping $X_{p^{(3,2,2)}}$ is pictured on the right.
  \vspace{-4ex}
  \[
\begin{tikzpicture}[line join=round,opacity=0.7]
\draw[arrows=->](.786,-.786)--(.561,-.354);
\filldraw[fill=gray](1.416,-.263)--(1.437,.072)--(1.333,.192)--(1.314,-.133)--cycle;
\filldraw[fill=gray](1.068,-.283)--(1.416,-.263)--(1.314,-.133)--(.978,-.151)--cycle;
\filldraw[fill=gray](.978,-.151)--(1.314,-.133)--(1.333,.192)--(.993,.175)--cycle;
\filldraw[fill=gray](1.068,-.283)--(1.084,.055)--(.993,.175)--(.978,-.151)--cycle;
\filldraw[fill=gray](1.084,.055)--(1.437,.072)--(1.333,.192)--(.993,.175)--cycle;
\filldraw[fill=gray](1.068,-.283)--(1.416,-.263)--(1.437,.072)--(1.084,.055)--cycle;
\draw[arrows=->](.786,-.786)--(1.855,-.718);
\draw[arrows=->](.786,-.786)--(.825,.267);
\draw (2.099,-.703) node {$t_0$};\draw (.832,.452) node {$t_1$};\draw (.435,-.115) node {$t_2$};\end{tikzpicture}
  \qquad\qquad\qquad
\begin{tikzpicture}[line join=round,opacity=0.7]
\draw[arrows=->](.786,-.786)--(.379,-.007);
\filldraw[fill=gray](2.517,.275)--(2.802,.288)--(2.838,.576)--(2.549,.564)--cycle;
\filldraw[fill=gray](2.647,.175)--(2.941,.19)--(2.802,.288)--(2.517,.275)--cycle;
\filldraw[fill=gray](2.941,.19)--(2.98,.487)--(2.838,.576)--(2.802,.288)--cycle;
\filldraw[fill=gray](2.647,.175)--(2.682,.473)--(2.549,.564)--(2.517,.275)--cycle;
\filldraw[fill=gray](2.682,.473)--(2.98,.487)--(2.838,.576)--(2.549,.564)--cycle;
\filldraw[fill=gray](2.05,.145)--(2.077,.446)--(1.963,.539)--(1.938,.247)--cycle;
\filldraw[fill=gray](1.746,.13)--(2.05,.145)--(1.938,.247)--(1.644,.233)--cycle;
\filldraw[fill=gray](1.644,.233)--(1.938,.247)--(1.963,.539)--(1.666,.526)--cycle;
\filldraw[fill=gray](1.746,.13)--(1.77,.432)--(1.666,.526)--(1.644,.233)--cycle;
\filldraw[fill=gray](1.77,.432)--(2.077,.446)--(1.963,.539)--(1.666,.526)--cycle;
\filldraw[fill=gray](1.131,.1)--(1.146,.404)--(1.062,.5)--(1.049,.204)--cycle;
\filldraw[fill=gray](.818,.084)--(1.131,.1)--(1.049,.204)--(.746,.19)--cycle;
\filldraw[fill=gray](.746,.19)--(1.049,.204)--(1.062,.5)--(.756,.487)--cycle;
\filldraw[fill=gray](2.647,.175)--(2.941,.19)--(2.98,.487)--(2.682,.473)--cycle;
\filldraw[fill=gray](.818,.084)--(.829,.39)--(.756,.487)--(.746,.19)--cycle;
\filldraw[fill=gray](2.616,1.165)--(2.912,1.175)--(2.95,1.487)--(2.651,1.478)--cycle;
\filldraw[fill=gray](2.755,1.094)--(3.061,1.104)--(2.912,1.175)--(2.616,1.165)--cycle;
\filldraw[fill=gray](3.061,1.104)--(3.103,1.426)--(2.95,1.487)--(2.912,1.175)--cycle;
\filldraw[fill=gray](.829,.39)--(1.146,.404)--(1.062,.5)--(.756,.487)--cycle;
\filldraw[fill=gray](2.755,1.094)--(2.793,1.417)--(2.651,1.478)--(2.616,1.165)--cycle;
\filldraw[fill=gray](1.746,.13)--(2.05,.145)--(2.077,.446)--(1.77,.432)--cycle;
\filldraw[fill=gray](2.793,1.417)--(3.103,1.426)--(2.95,1.487)--(2.651,1.478)--cycle;
\filldraw[fill=gray](2.134,1.072)--(2.164,1.398)--(2.042,1.461)--(2.015,1.145)--cycle;
\filldraw[fill=gray](1.71,1.135)--(2.015,1.145)--(2.042,1.461)--(1.733,1.452)--cycle;
\filldraw[fill=gray](1.819,1.061)--(2.134,1.072)--(2.015,1.145)--(1.71,1.135)--cycle;
\filldraw[fill=gray](1.819,1.061)--(1.844,1.389)--(1.733,1.452)--(1.71,1.135)--cycle;
\filldraw[fill=gray](.818,.084)--(1.131,.1)--(1.146,.404)--(.829,.39)--cycle;
\filldraw[fill=gray](1.844,1.389)--(2.164,1.398)--(2.042,1.461)--(1.733,1.452)--cycle;
\filldraw[fill=gray](1.178,1.039)--(1.195,1.37)--(1.106,1.435)--(1.091,1.115)--cycle;
\filldraw[fill=gray](.777,1.104)--(1.091,1.115)--(1.106,1.435)--(.787,1.426)--cycle;
\filldraw[fill=gray](.853,1.028)--(1.178,1.039)--(1.091,1.115)--(.777,1.104)--cycle;
\filldraw[fill=gray](.853,1.028)--(.865,1.36)--(.787,1.426)--(.777,1.104)--cycle;
\filldraw[fill=gray](2.755,1.094)--(3.061,1.104)--(3.103,1.426)--(2.793,1.417)--cycle;
\filldraw[fill=gray](.865,1.36)--(1.195,1.37)--(1.106,1.435)--(.787,1.426)--cycle;
\filldraw[fill=gray](1.819,1.061)--(2.134,1.072)--(2.164,1.398)--(1.844,1.389)--cycle;
\filldraw[fill=gray](.853,1.028)--(1.178,1.039)--(1.195,1.37)--(.865,1.36)--cycle;
\filldraw[fill=gray](2.937,-.046)--(3.251,-.029)--(3.297,.287)--(2.979,.271)--cycle;
\filldraw[fill=gray](3.098,-.169)--(3.424,-.151)--(3.251,-.029)--(2.937,-.046)--cycle;
\filldraw[fill=gray](3.424,-.151)--(3.474,.175)--(3.297,.287)--(3.251,-.029)--cycle;
\filldraw[fill=gray](3.098,-.169)--(3.144,.158)--(2.979,.271)--(2.937,-.046)--cycle;
\filldraw[fill=gray](3.144,.158)--(3.474,.175)--(3.297,.287)--(2.979,.271)--cycle;
\filldraw[fill=gray](2.436,-.206)--(2.473,.125)--(2.331,.24)--(2.298,-.08)--cycle;
\filldraw[fill=gray](2.1,-.225)--(2.436,-.206)--(2.298,-.08)--(1.973,-.098)--cycle;
\filldraw[fill=gray](1.973,-.098)--(2.298,-.08)--(2.331,.24)--(2.002,.224)--cycle;
\filldraw[fill=gray](2.1,-.225)--(2.132,.107)--(2.002,.224)--(1.973,-.098)--cycle;
\filldraw[fill=gray](2.132,.107)--(2.473,.125)--(2.331,.24)--(2.002,.224)--cycle;
\filldraw[fill=gray](1.416,-.263)--(1.437,.072)--(1.333,.192)--(1.314,-.133)--cycle;
\filldraw[fill=gray](1.068,-.283)--(1.416,-.263)--(1.314,-.133)--(.978,-.151)--cycle;
\filldraw[fill=gray](.978,-.151)--(1.314,-.133)--(1.333,.192)--(.993,.175)--cycle;
\filldraw[fill=gray](3.098,-.169)--(3.424,-.151)--(3.474,.175)--(3.144,.158)--cycle;
\filldraw[fill=gray](1.068,-.283)--(1.084,.055)--(.993,.175)--(.978,-.151)--cycle;
\filldraw[fill=gray](3.066,.934)--(3.393,.946)--(3.444,1.29)--(3.112,1.279)--cycle;
\filldraw[fill=gray](3.24,.844)--(3.58,.857)--(3.393,.946)--(3.066,.934)--cycle;
\filldraw[fill=gray](3.58,.857)--(3.635,1.214)--(3.444,1.29)--(3.393,.946)--cycle;
\filldraw[fill=gray](1.084,.055)--(1.437,.072)--(1.333,.192)--(.993,.175)--cycle;
\filldraw[fill=gray](3.24,.844)--(3.29,1.202)--(3.112,1.279)--(3.066,.934)--cycle;
\filldraw[fill=gray](2.1,-.225)--(2.436,-.206)--(2.473,.125)--(2.132,.107)--cycle;
\draw[arrows=->](.786,-.786)--(3.886,-.589);
\filldraw[fill=gray](3.29,1.202)--(3.635,1.214)--(3.444,1.29)--(3.112,1.279)--cycle;
\filldraw[fill=gray](2.549,.817)--(2.589,1.179)--(2.436,1.258)--(2.4,.909)--cycle;
\filldraw[fill=gray](2.061,.896)--(2.4,.909)--(2.436,1.258)--(2.093,1.247)--cycle;
\filldraw[fill=gray](2.198,.803)--(2.549,.817)--(2.4,.909)--(2.061,.896)--cycle;
\filldraw[fill=gray](2.198,.803)--(2.232,1.167)--(2.093,1.247)--(2.061,.896)--cycle;
\filldraw[fill=gray](1.068,-.283)--(1.416,-.263)--(1.437,.072)--(1.084,.055)--cycle;
\filldraw[fill=gray](2.232,1.167)--(2.589,1.179)--(2.436,1.258)--(2.093,1.247)--cycle;
\filldraw[fill=gray](1.118,.761)--(1.482,.775)--(1.373,.87)--(1.023,.857)--cycle;
\filldraw[fill=gray](1.482,.775)--(1.506,1.143)--(1.394,1.225)--(1.373,.87)--cycle;
\filldraw[fill=gray](1.023,.857)--(1.373,.87)--(1.394,1.225)--(1.038,1.214)--cycle;
\filldraw[fill=gray](1.118,.761)--(1.136,1.131)--(1.038,1.214)--(1.023,.857)--cycle;
\filldraw[fill=gray](3.24,.844)--(3.58,.857)--(3.635,1.214)--(3.29,1.202)--cycle;
\filldraw[fill=gray](1.136,1.131)--(1.506,1.143)--(1.394,1.225)--(1.038,1.214)--cycle;
\filldraw[fill=gray](2.198,.803)--(2.549,.817)--(2.589,1.179)--(2.232,1.167)--cycle;
\draw[arrows=->](.786,-.786)--(.867,1.428);
\filldraw[fill=gray](1.118,.761)--(1.482,.775)--(1.506,1.143)--(1.136,1.131)--cycle;
\draw (4.114,-.574) node {$t_0$};\draw (.875,1.633) node {$t_1$};\draw (.277,.188) node {$t_2$};\end{tikzpicture}
  \]
\end{example}

Given two spaces~$X$ and~$Y$ which are hypercubes with holes (which is the case
for the geometric realizations of the programs we are considering here), we
write $X\glue jY$ for the space obtained by identifying the $j$-th target face
of the hypercube~$X$ with the $j$\nbd{}th source face of the hypercube~$Y$, and
call it the \emph{$j$\nbd{}gluing} of~$X$ and~$Y$. Formally, this can be defined
as in Section~\ref{geomsem} as $X\glue jY=X\uplus Y/\sim$, where the
relation~$\sim$ identifies points~$x\in X$ and~$y\in Y$ such that $x_j=\infty$,
$y_j=0$ and~$x_{j'}=y_{j'}$ for every dimension~$j'\neq j$, and directed paths
are defined in a similar fashion. Notice that, by definition, there is a
canonical embedding of $X$ (\resp $Y$) into $X\glue jY$, which will allow us to
implicitly consider~$X$ (\resp $Y$) as a subspace of $X\glue jY$ in the
following.

\begin{example}
  \label{ex:glue322}
  The $(3,2,2)$-delooping of Example~\ref{ex:delooping} is
  \[
  X_{p^{(3,2,2)}}
  \qeq
  (Y\glue 1 Y)\glue 2(Y\glue 1 Y)
  \qquad
  \text{with}
  \qquad
  Y=X_p\glue 0X_p\glue 0X_p
  \]
\end{example}

\noindent
More generally, any $v$\nbd{}delooping~$p^v$ of a program~$p$ of the
form~\eqref{eq:prog} can be obtained by gluing copies $X_p^w$ of~$X_p$, indexed
by a vector $w$ such that for every dimension~$i$ with $0\leq i<n$, we have
$0\leq w_i<v_i$ (what we will simply write $0\leq w<v$).

\newcommand{\shadow}[2]{#2|_{#1}}

Given two scheduling matrices~$M$ and~$N$ encoding extensions of holes of such a
program~$p$ (\cf Section~\ref{indexposet}), we reuse the notation and
write~$M\glue j N$ for the obvious matrix coding extension of holes in the
space~$X_p\glue j X_p$. At this point, it is crucial to notice that the holes
described by~$N$ in the second copy of~$X_p$ can have an effect on the first
copy of~$X_p$ (when they are extended to~$0$ in the direction~$j$), what we call
the \emph{$j$-shadow of~$N$}, and write $X_{\shadow jN}$.

\begin{example}
  With the program~$p$ of Example~\ref{ex:delooping}, consider the matrices
  $M=(\begin{matrix}1&0&0\end{matrix})$ and
  $N=(\begin{matrix}0&0&1\end{matrix})$. We have $M\glue
  0N=\left(\begin{matrix}1&0&0\\0&0&1\end{matrix}\right)$, the space $X_{M\glue
    0 N}$ is pictured on the left, and the $0$-shadow $X_{\shadow 0N}$ of~$N$ is
  pictured on the right:
  \[
\begin{tikzpicture}[line join=round,opacity=0.7]
\draw[arrows=->](.786,-.786)--(.561,-.354);
\filldraw[fill=gray](2.436,-.206)--(2.473,.125)--(2.331,.24)--(2.298,-.08)--cycle;
\filldraw[fill=lightgray](2.401,-.527)--(2.473,.125)--(2.331,.24)--(2.266,-.391)--cycle;
\filldraw[fill=gray](1.973,-.098)--(2.298,-.08)--(2.331,.24)--(2.002,.224)--cycle;
\filldraw[fill=lightgray](.63,-.487)--(1.295,-.448)--(1.333,.192)--(.649,.158)--cycle;
\filldraw[fill=lightgray](.63,-.487)--(2.266,-.391)--(2.331,.24)--(.649,.158)--cycle;
\filldraw[fill=lightgray](.705,-.631)--(2.401,-.527)--(2.266,-.391)--(.63,-.487)--cycle;
\filldraw[fill=gray](2.1,-.225)--(2.436,-.206)--(2.298,-.08)--(1.973,-.098)--cycle;
\filldraw[fill=gray](1.416,-.263)--(1.437,.072)--(1.333,.192)--(1.314,-.133)--cycle;
\filldraw[fill=gray](2.1,-.225)--(2.132,.107)--(2.002,.224)--(1.973,-.098)--cycle;
\filldraw[fill=lightgray,draw=none](1.395,-.589)--(2.401,-.527)--(2.473,.125)--(1.437,.072)--cycle;
\draw(1.395,-.589)--(2.401,-.527)--(2.473,.125)--(1.437,.072);
\filldraw[fill=lightgray](1.503,-.741)--(1.55,-.057)--(1.333,.192)--(1.295,-.448)--cycle;
\filldraw[fill=lightgray](1.147,-.763)--(1.503,-.741)--(1.295,-.448)--(.964,-.468)--cycle;
\filldraw[fill=lightgray](.964,-.468)--(1.295,-.448)--(1.333,.192)--(.993,.175)--cycle;
\filldraw[fill=lightgray](1.395,-.589)--(1.437,.072)--(1.333,.192)--(1.295,-.448)--cycle;
\filldraw[fill=lightgray](.705,-.631)--(1.395,-.589)--(1.295,-.448)--(.63,-.487)--cycle;
\filldraw[fill=gray](1.068,-.283)--(1.416,-.263)--(1.314,-.133)--(.978,-.151)--cycle;
\filldraw[fill=gray](.978,-.151)--(1.314,-.133)--(1.333,.192)--(.993,.175)--cycle;
\filldraw[fill=lightgray,draw=none](1.052,-.61)--(1.395,-.589)--(1.437,.072)--(1.084,.055)--cycle;
\draw(1.052,-.61)--(1.395,-.589)--(1.437,.072)--(1.084,.055);
\filldraw[fill=lightgray,draw=none](1.052,-.61)--(1.395,-.589)--(1.437,.072)--(1.084,.055)--cycle;
\draw(1.052,-.61)--(1.395,-.589);
\draw(1.437,.072)--(1.084,.055);
\filldraw[fill=gray](1.068,-.283)--(1.416,-.263)--(1.437,.072)--(1.084,.055)--cycle;
\filldraw[fill=lightgray](1.147,-.763)--(1.183,-.076)--(.993,.175)--(.964,-.468)--cycle;
\filldraw[fill=lightgray](.705,-.631)--(.727,.037)--(.649,.158)--(.63,-.487)--cycle;
\filldraw[fill=lightgray](.705,-.631)--(.727,.037)--(.649,.158)--(.63,-.487)--cycle;
\filldraw[fill=gray](1.068,-.283)--(1.084,.055)--(.993,.175)--(.978,-.151)--cycle;
\filldraw[fill=lightgray](.727,.037)--(2.473,.125)--(2.331,.24)--(.649,.158)--cycle;
\filldraw[fill=gray](2.132,.107)--(2.473,.125)--(2.331,.24)--(2.002,.224)--cycle;
\filldraw[fill=lightgray](.727,.037)--(1.437,.072)--(1.333,.192)--(.649,.158)--cycle;
\filldraw[fill=gray](1.084,.055)--(1.437,.072)--(1.333,.192)--(.993,.175)--cycle;
\filldraw[fill=lightgray](1.183,-.076)--(1.55,-.057)--(1.333,.192)--(.993,.175)--cycle;
\filldraw[fill=gray](2.1,-.225)--(2.436,-.206)--(2.473,.125)--(2.132,.107)--cycle;
\filldraw[fill=lightgray,draw=none](.705,-.631)--(1.052,-.61)--(1.084,.055)--(.727,.037)--cycle;
\draw(1.084,.055)--(.727,.037)--(.705,-.631)--(1.052,-.61);
\filldraw[fill=lightgray,draw=none](.705,-.631)--(1.052,-.61)--(1.084,.055)--(.727,.037)--cycle;
\draw(1.084,.055)--(.727,.037)--(.705,-.631)--(1.052,-.61);
\draw[arrows=->](.786,-.786)--(2.887,-.652);
\filldraw[fill=lightgray](1.147,-.763)--(1.503,-.741)--(1.55,-.057)--(1.183,-.076)--cycle;
\draw[arrows=->](.786,-.786)--(.825,.267);
\draw (3.123,-.637) node {$t_0$};\draw (.832,.452) node {$t_1$};\draw (.435,-.115) node {$t_2$};\end{tikzpicture}
  \qquad\qquad
\begin{tikzpicture}[line join=round,opacity=0.7]
\draw[arrows=->](.786,-.786)--(.561,-.354);
\filldraw[fill=lightgray](1.734,-.568)--(1.786,.09)--(1.669,.208)--(1.622,-.429)--cycle;
\filldraw[fill=lightgray](.705,-.631)--(1.734,-.568)--(1.622,-.429)--(.63,-.487)--cycle;
\filldraw[fill=lightgray](.63,-.487)--(1.622,-.429)--(1.669,.208)--(.649,.158)--cycle;
\filldraw[fill=lightgray](.705,-.631)--(.727,.037)--(.649,.158)--(.63,-.487)--cycle;
\filldraw[fill=gray](.978,-.151)--(1.314,-.133)--(1.333,.192)--(.993,.175)--cycle;
\filldraw[fill=gray](1.068,-.283)--(1.416,-.263)--(1.314,-.133)--(.978,-.151)--cycle;
\filldraw[fill=gray](1.416,-.263)--(1.437,.072)--(1.333,.192)--(1.314,-.133)--cycle;
\filldraw[fill=gray](1.068,-.283)--(1.084,.055)--(.993,.175)--(.978,-.151)--cycle;
\filldraw[fill=lightgray](.727,.037)--(1.786,.09)--(1.669,.208)--(.649,.158)--cycle;
\filldraw[fill=gray](1.084,.055)--(1.437,.072)--(1.333,.192)--(.993,.175)--cycle;
\filldraw[fill=lightgray](.705,-.631)--(1.734,-.568)--(1.786,.09)--(.727,.037)--cycle;
\filldraw[fill=gray](1.068,-.283)--(1.416,-.263)--(1.437,.072)--(1.084,.055)--cycle;
\draw[arrows=->](.786,-.786)--(1.855,-.718);
\draw[arrows=->](.786,-.786)--(.825,.267);
\draw (2.099,-.703) node {$t_0$};\draw (.832,.452) node {$t_1$};\draw (.435,-.115) node {$t_2$};\end{tikzpicture}
  \]
\end{example}

The above example makes clear that the space corresponding to a scheduling
$M\glue j N$ is of the form $X_{M\glue jN}=(X_M\cap X_{\shadow jN})\otimes_j
X_N$, \ie the holes in the first copy come either from~$M$ or from shadows
of~$N$. Moreover, the holes in the space $X_{\shadow jN}$ are hypercubes which
are products of intervals of the form $\prod_{0\leq j<n}\dui_j$, where each
interval~$\dui_j$ is of the form $]x_j^i,y_j^i[$ or $[0,y_j^i[$ or $[0,\infty]$,
with $0\leq i<l$. The shadows can therefore be coded as matrices (using a
slightly different coding from the one used up to now, the precise way they are
coded being quite irrelevant) and we write $\shadow jN$ for the matrix coding
the $j$-shadow of~$n$, which can easily be computed from~$N$ and~$j$. A
scheduling matrix $M$ can obviously be seen as a particular ``shadow'', enabling
us to use the same notation for both, and we write $M\cup N$ for the union of
two shadows $M$ and $N$, so that $X_{M\cup N}=X_M\cap X_N$. Finally, given a
shadow~$M$, the algorithm described in Section~\ref{sec:homotopy-classes} can
easily be adapted to the new coding in order to determine whether the
space~$X_M$ is alive.

\subsection{The shadow automaton}
The trace space of a program $p^*$ is not finite in the general case. We show
here that it can however be described as the set of paths of an automaton that
we call the \emph{shadow automaton}: this automaton provides us with a
\emph{finite presentation} of the set of schedulings.

Consider the $v$-delooping~$p^v$ of a program~$p$. The space~$X_{p^v}$ consists
of the gluing of copies of~$X_p$ indexed by vectors~$w$ such that $0\leq w<v$
and similarly, a scheduling~$M$ of~$X_{p^v}$ consists of the gluing of matrices
$M^w$. Clearly, if some submatrix~$M^w$ is dead then the whole matrix~$M$ is
dead:

\begin{lemma}
  If a matrix~$M$ is alive then all its submatrices~$M^w$ are alive.
\end{lemma}

\noindent
However, the converse is not true because a scheduling~$M^w$ might create a
deadlock with the shadows coming from matrices above it. For instance in
Example~\ref{ex:delooping}, the matrix $M=(\begin{matrix}1&0&0\end{matrix})\glue
0(\begin{matrix}0&1&1\end{matrix})$ is not alive because the
space~$X_{M^{(0,0,0)}}$ induced by the submatrix $M^{(0,0,0)}$ is contained in
the space~$X_N$, where $N=(\begin{matrix}1&1&1\end{matrix})$ is a dead matrix:
\vspace{-1ex}
\[
\begin{tikzpicture}[line join=round,opacity=0.7]
\draw[arrows=->](.786,-.786)--(.561,-.354);
\filldraw[fill=lightgray](2.586,-.342)--(2.626,0)--(2.331,.24)--(2.298,-.08)--cycle;
\filldraw[fill=gray](2.436,-.206)--(2.473,.125)--(2.331,.24)--(2.298,-.08)--cycle;
\filldraw[fill=lightgray](2.401,-.527)--(2.473,.125)--(2.331,.24)--(2.266,-.391)--cycle;
\filldraw[fill=gray](1.973,-.098)--(2.298,-.08)--(2.331,.24)--(2.002,.224)--cycle;
\filldraw[fill=lightgray](.63,-.487)--(2.266,-.391)--(2.331,.24)--(.649,.158)--cycle;
\filldraw[fill=lightgray](.705,-.631)--(2.401,-.527)--(2.266,-.391)--(.63,-.487)--cycle;
\filldraw[fill=lightgray](.639,-.169)--(2.298,-.08)--(2.331,.24)--(.649,.158)--cycle;
\filldraw[fill=gray](2.1,-.225)--(2.436,-.206)--(2.298,-.08)--(1.973,-.098)--cycle;
\filldraw[fill=lightgray,draw=none](1.395,-.589)--(1.416,-.263)--(1.314,-.133)--(1.295,-.448)--cycle;
\draw(1.314,-.133)--(1.295,-.448)--(1.395,-.589);
\filldraw[fill=lightgray](.964,-.468)--(1.295,-.448)--(1.333,.192)--(.993,.175)--cycle;
\filldraw[fill=lightgray](1.147,-.763)--(1.503,-.741)--(1.295,-.448)--(.964,-.468)--cycle;
\filldraw[fill=lightgray,draw=none](1.052,-.61)--(1.068,-.283)--(.978,-.151)--(.964,-.468)--cycle;
\draw(.978,-.151)--(.964,-.468)--(1.052,-.61);
\filldraw[fill=lightgray,draw=none](1.416,-.263)--(1.395,-.589)--(2.401,-.527)--(2.436,-.206)--cycle;
\draw(1.395,-.589)--(2.401,-.527)--(2.436,-.206);
\filldraw[fill=lightgray,draw=none](1.395,-.589)--(1.503,-.741)--(1.526,-.404)--(1.416,-.263)--cycle;
\draw(1.395,-.589)--(1.503,-.741)--(1.526,-.404);
\filldraw[fill=lightgray,draw=none](1.416,-.263)--(1.068,-.283)--(1.052,-.61)--(1.395,-.589)--cycle;
\draw(1.052,-.61)--(1.395,-.589);
\filldraw[fill=lightgray,draw=none](1.052,-.61)--(1.147,-.763)--(1.165,-.425)--(1.068,-.283)--cycle;
\draw(1.052,-.61)--(1.147,-.763)--(1.165,-.425);
\filldraw[fill=lightgray,draw=none](1.068,-.283)--(.716,-.302)--(.705,-.631)--(1.052,-.61)--cycle;
\draw(.716,-.302)--(.705,-.631)--(1.052,-.61);
\filldraw[fill=lightgray](.799,-.446)--(2.586,-.342)--(2.298,-.08)--(.639,-.169)--cycle;
\filldraw[fill=gray](2.1,-.225)--(2.132,.107)--(2.002,.224)--(1.973,-.098)--cycle;
\filldraw[fill=lightgray](.705,-.631)--(.727,.037)--(.649,.158)--(.63,-.487)--cycle;
\filldraw[fill=lightgray,draw=none](1.416,-.263)--(1.437,.072)--(1.333,.192)--(1.314,-.133)--cycle;
\draw(1.437,.072)--(1.333,.192)--(1.314,-.133);
\filldraw[fill=gray](.978,-.151)--(1.314,-.133)--(1.333,.192)--(.993,.175)--cycle;
\filldraw[fill=gray](1.416,-.263)--(1.437,.072)--(1.333,.192)--(1.314,-.133)--cycle;
\filldraw[fill=gray](1.068,-.283)--(1.416,-.263)--(1.314,-.133)--(.978,-.151)--cycle;
\filldraw[fill=lightgray,draw=none](1.068,-.283)--(1.084,.055)--(.993,.175)--(.978,-.151)--cycle;
\draw(1.084,.055)--(.993,.175)--(.978,-.151);
\filldraw[fill=gray](1.068,-.283)--(1.084,.055)--(.993,.175)--(.978,-.151)--cycle;
\filldraw[fill=lightgray](.727,.037)--(2.473,.125)--(2.331,.24)--(.649,.158)--cycle;
\filldraw[fill=lightgray,draw=none](.716,-.302)--(2.436,-.206)--(2.473,.125)--(.727,.037)--cycle;
\draw(2.436,-.206)--(2.473,.125)--(.727,.037)--(.716,-.302);
\filldraw[fill=gray](2.1,-.225)--(2.436,-.206)--(2.473,.125)--(2.132,.107)--cycle;
\filldraw[fill=lightgray,draw=none](1.416,-.263)--(1.526,-.404)--(1.55,-.057)--(1.437,.072)--cycle;
\draw(1.526,-.404)--(1.55,-.057)--(1.437,.072);
\filldraw[fill=gray](1.068,-.283)--(1.416,-.263)--(1.437,.072)--(1.084,.055)--cycle;
\filldraw[fill=lightgray,draw=none](1.068,-.283)--(1.165,-.425)--(1.183,-.076)--(1.084,.055)--cycle;
\draw(1.165,-.425)--(1.183,-.076)--(1.084,.055);
\filldraw[fill=lightgray](.812,-.095)--(2.626,0)--(2.331,.24)--(.649,.158)--cycle;
\filldraw[fill=gray](2.132,.107)--(2.473,.125)--(2.331,.24)--(2.002,.224)--cycle;
\filldraw[fill=lightgray](.799,-.446)--(.812,-.095)--(.649,.158)--(.639,-.169)--cycle;
\filldraw[fill=gray](1.084,.055)--(1.437,.072)--(1.333,.192)--(.993,.175)--cycle;
\filldraw[fill=lightgray](1.183,-.076)--(1.55,-.057)--(1.333,.192)--(.993,.175)--cycle;
\draw[arrows=->](.786,-.786)--(2.887,-.652);
\filldraw[fill=lightgray](.799,-.446)--(2.586,-.342)--(2.626,0)--(.812,-.095)--cycle;
\filldraw[fill=lightgray](1.147,-.763)--(1.503,-.741)--(1.55,-.057)--(1.183,-.076)--cycle;
\draw[arrows=-](.786,-.786)--(.799,-.446);
\draw[arrows=-](.799,-.446)--(.812,-.095);
\draw[arrows=->](.812,-.095)--(.825,.267);
\draw (3.123,-.637) node {$t_0$};\draw (.832,.452) node {$t_1$};\draw (.435,-.115) node {$t_2$};\end{tikzpicture}
\]

In order to generate all the possible schedulings $M^w$ visited by a total path
in~$X_{p^v}$, we therefore have to take in account the shadows dropped by
scheduling of copies of~$X_p$ in its future. We will construct an automaton
which will consider the visited schedulings of the path, starting from the end,
and maintains the shadow they produce on the next state in a given
direction~$j$, so that we can compute the possible previous matrices in
direction~$j$ such that the whole matrix is not dead. Formally,

\begin{definition}
  \label{shadow-automaton}
  The \emph{shadow automaton} of a program~$p$ is a non-deterministic automaton
  whose
  \begin{itemize}
  \item states are shadows
  \item transitions $\vxym{N\ar[r]^{j,M}&N'}$ are labeled by a direction $j$
    (with $0\leq j<n$) and a scheduling~$M$
  \end{itemize}
  defined as the smallest automaton
  \begin{itemize}
  \item containing the empty scheduling~$\emptyset$
  \item and such that for every state $N'$, for every direction~$j$ and for
    every scheduling~$M$ such that the scheduling $M\cup N'$ is alive, and $M$
    is maximal with this property, there is a transition
    $\vxym{N\ar[r]^{j,M}&N'}$ with~$N=\shadow j{(M\cup N')}$.
  \end{itemize}
  All the states of the automaton are both initial and final.
\end{definition}

\begin{example}
  \label{ex:shadow-automaton}
  Consider the program $p=q|q$ with $q=\P a.\V a$ whose geometric semantics is a
  square with a square hole. The associated shadow automaton is
  \vspace{-2ex}
  \[
  \xymatrix@C=12ex@R=6ex{
    \ar@(ul,dl)_-{1,\includegraphics[scale=0.15]{sa_m0}}{\includegraphics[scale=0.15]{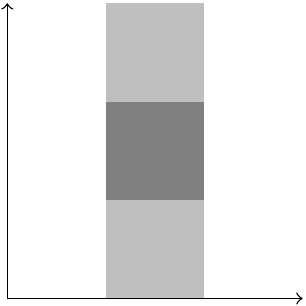}}\ar@/^/[r]^-{1,\includegraphics[scale=0.15]{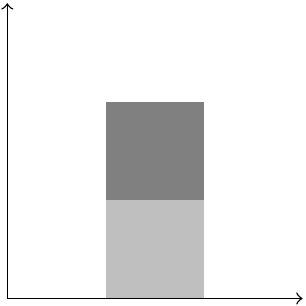}}&\ar@(ul,ur)^-{0,\includegraphics[scale=0.15]{sa_m0}}\ar@(dl,dr)_-{1,\includegraphics[scale=0.15]{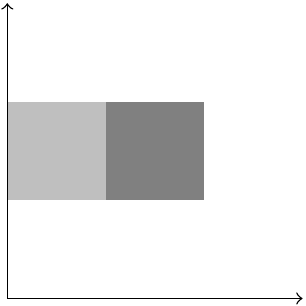}}{\includegraphics[scale=0.15]{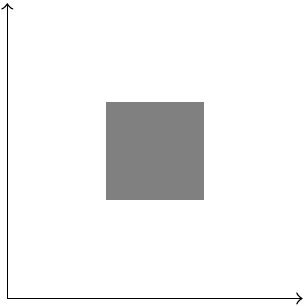}}\ar@/^/[r]^-{1,\includegraphics[scale=0.15]{sa_m1}}\ar@/^/[l]^-{0,\includegraphics[scale=0.15]{sa_m0}}&\ar@/^/[l]^-{0,\includegraphics[scale=0.15]{sa_m1}}{\includegraphics[scale=0.15]{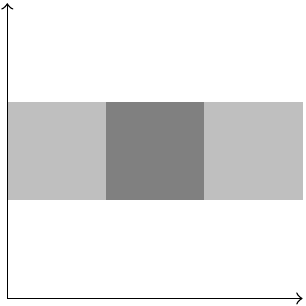}}\ar@(ur,dr)^-{0,\includegraphics[scale=0.15]{sa_m1}}\\
    &\\
  }
  \]

  \vspace{-4ex}\noindent For instance the transition
  $\xymatrix{{\includegraphics[scale=0.15]{sa_s1}}\ar[r]^-{0,\includegraphics[scale=0.15]{sa_m1}}&{\includegraphics[scale=0.15]{sa_semp}}}$
  is computed as follows: we take the shadow
  $M=\includegraphics[scale=0.15]{sa_m1}\cup\includegraphics[scale=0.15]{sa_semp}=\includegraphics[scale=0.15]{sa_m1}$
  and compute its shadow in direction $0$, \ie on the left, to compute the
  source of the transition. This shadow is
  $\includegraphics[scale=0.15]{sa_s1}$, namely:
  $\includegraphics[scale=0.15]{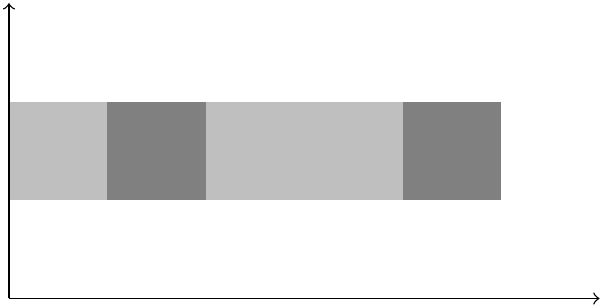}$.
\end{example}

The interest of the automaton lies in the fact that fully describes the possible
schedulings crossed by a total path in a scheduling of a delooping $X_{p^v}$:
\begin{theorem}
  \label{thm:sa-path-lift}
  Suppose that~$M$ is a scheduling of $X_{p^v}$, obtained by gluing schedulings
  $M^w$ of $X_p$. Then there exists a total path in~$X_M$ going through the
  subspaces $X_{M^{w_0}},X_{M^{w_1}},\ldots,X_{M^{w_m}}$ in this order, such
  that $w_k$ and $w_{k+1}$ only differ by one coordinate~$j_k$ (\ie the path
  exits from $X_{M^{w_k}}$ through its $j_k$\nbd{}th face),
  if and only if there exists a path labeled as follows in the shadow automaton:
  \[
  \xymatrix@C=10ex{
    N_0\ar[r]^-{j,M^{w_0}}&N_1\ar[r]^-{j_0,M^{w_1}}&N_2\ar@{}[r]|-{\ldots}&N_m\ar[r]^-{j_{m-1},M^{w_m}}&N_{m+1}\\
  }
  \]
  for some states $N_i$ and dimension~$j$.
\end{theorem}

\begin{example}
  \label{ex:paths}
  With the program~$p$ of Example~\ref{ex:shadow-automaton}, the following paths
  in the $(2,2)$\nbd{}delooping
  \vspace{-2ex}
  \[
  \begin{tikzpicture}[scale=.4]
    \fill[fill=lightgray] (1,0) -- (1,4) -- (2,4) -- (2,0) -- cycle;
    \fill[fill=lightgray] (3,0) -- (3,4) -- (4,4) -- (4,0) -- cycle;
    \draw[line width=0.3mm,color=red] (0,0) .. controls (0.4,4.8) .. (5,5);
    \fill[fill=gray] (1,1) -- (2,1) -- (2,2) -- (1,2) -- cycle;
    \fill[fill=gray] (3,1) -- (4,1) -- (4,2) -- (3,2) -- cycle;
    \fill[fill=gray] (1,3) -- (2,3) -- (2,4) -- (1,4) -- cycle;
    \fill[fill=gray] (3,3) -- (4,3) -- (4,4) -- (3,4) -- cycle;
    \draw[style=dotted] (0,2.5) -- (5,2.5);
    \draw[style=dotted] (2.5,0) -- (2.5,5);
    \draw[->] (0,0) -- (5,0);
    \draw[->] (0,0) -- (0,5);
    \draw (5.5,0) node {$t_0$};
    \draw (0,5.5) node {$t_1$};
  \end{tikzpicture}
  \qquad\qquad
  \begin{tikzpicture}[scale=.4]
    \fill[fill=lightgray] (1,0) -- (1,2) -- (2,2) -- (2,0) -- cycle;
    \fill[fill=lightgray] (0,3) -- (0,4) -- (4,4) -- (4,3) -- cycle;
    \fill[fill=lightgray] (3,0) -- (3,2) -- (4,2) -- (4,0) -- cycle;
    \draw[line width=0.3mm,color=red] (0,0) .. controls (0.4,2.3) .. (2.5,2.3) .. controls (4.7,2.5) .. (5,5);
    \fill[fill=gray] (1,1) -- (2,1) -- (2,2) -- (1,2) -- cycle;
    \fill[fill=gray] (3,1) -- (4,1) -- (4,2) -- (3,2) -- cycle;
    \fill[fill=gray] (1,3) -- (2,3) -- (2,4) -- (1,4) -- cycle;
    \fill[fill=gray] (3,3) -- (4,3) -- (4,4) -- (3,4) -- cycle;
    \draw[style=dotted] (0,2.5) -- (5,2.5);
    \draw[style=dotted] (2.5,0) -- (2.5,5);
    \draw[->] (0,0) -- (5,0);
    \draw[->] (0,0) -- (0,5);
    \draw (5.5,0) node {$t_0$};
    \draw (0,5.5) node {$t_1$};
  \end{tikzpicture}
  \]
  are respectively witnessed by the following paths of the shadow automaton:
  \[
  \begin{array}{c}
    \vxym{
      {\includegraphics[scale=0.15]{sa_s0}}\ar[r]^-{1,\includegraphics[scale=0.15]{sa_m0}}&{\includegraphics[scale=0.15]{sa_s0}}\ar[r]^-{1,\includegraphics[scale=0.15]{sa_m0}}&{\includegraphics[scale=0.15]{sa_semp}}\ar[r]^-{0,\includegraphics[scale=0.15]{sa_m0}}&{\includegraphics[scale=0.15]{sa_semp}}
    }
    \qquad\qquad
    \vxym{
      {\includegraphics[scale=0.15]{sa_semp}}\ar[r]^-{0,\includegraphics[scale=0.15]{sa_m0}}&{\includegraphics[scale=0.15]{sa_semp}}\ar[r]^-{0,\includegraphics[scale=0.15]{sa_m0}}&{\includegraphics[scale=0.15]{sa_semp}}\ar[r]^-{1,\includegraphics[scale=0.15]{sa_m1}}&{\includegraphics[scale=0.15]{sa_semp}}
    }
  \end{array}
  \]
\end{example}

\subsection{Reducing the size of the shadow automaton}
The size of the shadow automaton grows very quickly when the complexity of the
trace space grows. For instance, for the program~$p$ of
Example~\ref{ex:delooping}, the shadow automaton has already 19 states and 80
transitions. We describe here some ways to reduce the automaton while preserving
Theorem~\ref{thm:sa-path-lift}. Namely, we should remark that the automaton is
not minimal in the following sense. By Proposition~\ref{prop:alive-dihom}, given
a scheduling~$M$ two total paths~$X_M$ are necessarily homotopic: an alive
scheduling thus describes an homotopy class of total paths. By
Theorem~\ref{thm:sa-path-lift}, the schedulings ``visited'' by a total path
in~$X_{p^v}$ are described by a path in the shadow automaton, therefore every
homotopy class of total paths in~$X_{p^v}$ is described by at least one path in
the scheduling automaton. The shadow automaton is not minimal in the sense that
generally, an homotopy class is described by more than one path in the
scheduling automaton.

\paragraph{Determinization.}
First, our non-deterministic automaton can be determinized using classical
algorithms of automata theory, which in practice greatly reduce their size: the
determinized automaton for the program of Example~\ref{ex:delooping} has only 4
states and 24 transitions.

\begin{example}
  \label{ex:sh-det}
  The determinized automata for Examples~\ref{ex:shadow-automaton}
  and~\ref{ex:delooping} are respectively:
  \[
  \xymatrix{
    &\ar@/_/[dl]_-{\_,\includegraphics[scale=0.15]{sa_m0}}I\ar@/^/[dr]^-{\_,\includegraphics[scale=0.15]{sa_m1}}&\\
    \ar@(ul,dl)_-{\_,\includegraphics[scale=0.15]{sa_m0}}\ar@/^/[rr]^-{1,\includegraphics[scale=0.15]{sa_m1}}0&&1\ar@/^/[ll]^-{0,\includegraphics[scale=0.15]{sa_m0}}\ar@(ur,dr)^-{\_,\includegraphics[scale=0.15]{sa_m1}}\\
  }
  \qquad\qquad
  \xymatrix {
    &\ar@/_/[ddl]_-{\_,M_1}I\ar[d]\ar@/^/[ddr]^-{\_,M_2}&\\
    &\ar@/^/[dl]|-{1,M_0}0\ar@(ul,ur)\ar@/^/[dr]|-{2,M_0}&\\
    1\ar@(ul,dl)_-{\_,M_1}\ar@/^/[ur]|-{0,M_1}\ar@/^/[rr]|-{2,M_1}&&2\ar@/^/[ll]|-{1,M_2}\ar@/^/[ul]|-{0,M_2}\ar@(ur,dr)^-{\_,M_2}\\
  }
  \]
  with
  \vspace{-2ex}
  \[
  M_0=\includegraphics[scale=0.6]{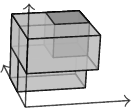}
  \qquad\qquad\qquad
  M_1=\includegraphics[scale=0.6]{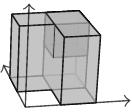}
  \qquad\qquad\qquad
  M_2=\includegraphics[scale=0.6]{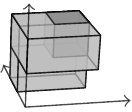}
  \]
  where ``$\_$'' means any direction $j$. The state $I$ is initial and all the
  states are final.
\end{example}

\paragraph{Quotient under connexity.}
A way to further reduce the automaton consists in quotienting the scheduling
matrices labeling the arrows of the automaton under the connexity relation of
Definition~\ref{def:connected} before determinizing the automaton, which is
formally justified by Proposition~\ref{prop:conn-comp}.

\begin{example}
  The shadow automaton corresponding to the program Example~\ref{ex:delooping}
  quotiented under connexity, determinized and minimized is simply the automaton
  $\vxym{ I\ar@(ur,dr)^{\_,M} }$ where $M=M_1=M_2=M_3$ up to connexity (the
  matrices $M_i$ are those defined in Example~\ref{ex:sh-det}).
\end{example}

We are currently investigating further conditions in order to construct the
minimal automaton describing the trace space associated to a looping program,
but the conditions mentioned above are already providing us with promisingly
small automata.

\subsection{Preliminary implementation and benchmark}
\label{sec:loops-implem}
A preliminary implementation of the computation of the shadow automaton was
done. The algorithm implemented is currently quite simple, but we plan to
generalize the algorithm of Section~\ref{sec:implem} soon, which is not
complicated from a theoretical point of view but much more involved technically,
in order to achieve better performances. Most experiments lead so far are
already promising and make it clear that taking in account the geometry of the
state-space enables us to reduce, sometimes drastically, the size of the control
flow graph corresponding to the program to be analyzed.

\begin{example}
  \label{ex:two-phase}
  The \emph{two-phase locking protocol} is a simple discipline for distributed
  databases, in which the processes first lock all the mutexes for the resources
  they are going to use and free all of them in the
  end~\cite{gunawardena1994homotopy}. This can be modeled as a program $q_{n,l}$
  consisting of $n$~copies of the process $p=\P {a_1}.\ldots\P{a_l}$
  $.\V{a_1}.\ldots\V{a_l}$ in parallel (each of these process is using $l$
  resources). For instance, the geometric semantics of $q_{2,2}=p|p$ is shown
  below. Notice that this state space is equivalent to a space with only one
  hole up to dihomotopy. More generally, given $l\geq 1$, it can be shown that
  the geometric semantics of $q_{n,l}$ is equivalent to $q_{n,1}$, which our
  algorithm is able to take into account! Namely, the size of the shadow
  automaton associated to $q_{n,l}^*$ only depends on~$n$ whereas the number of
  states of the automaton produced by SPIN is exponential in $l$ (with~$n$
  fixed). Below are presented the size (states, transitions) of the
  non-deterministic automaton ($s$, $t$), determinized automaton ($s'$,$t'$) and
  SPIN's automaton ($s_{\mathrm{SPIN}}$, $t_{\mathrm{SPIN}}$) for the two-phase
  locking process described in Example~\ref{ex:two-phase}, for some values
  of~$n$ and~$l$.
  \[
  \vcenter{
  \begin{tikzpicture}[scale=.3]
    \fill[fill=gray] (1,1) -- (3,1) -- (3,3) -- (1,3) -- cycle;
    \fill[fill=gray] (2,2) -- (4,2) -- (4,4) -- (2,4) -- cycle;
    \draw[->] (0,0) -- (5,0);
    \draw[->] (0,0) -- (0,5);
    \draw (5.5,0) node {$t_0$};
    \draw (0,5.5) node {$t_1$};
  \end{tikzpicture}}
  \hspace{-40ex}
  \begin{array}{r|r|r|r|r|r|r|r}
    n&l&s&t&s'&t'&s_{\mathrm{SPIN}}&t_{\mathrm{SPIN}}\\
    \hline{}
    2& 1& 3& 8& 3& 10&58&65\\
    2& 2& 3& 8& 3& 10&112&129\\
    2& 3& 3& 8& 3& 10&180&209\\
    3& 1& 19& 90& 4& 24&171&218\\
    3& 2& 19& 90& 4& 24&441&602\\
    3& 3& 19& 90& 4& 24&817&1128\\
  \end{array}
  \]
\end{example}

\subsection{An Application to static analysis}
\label{sec:static-anal}
Now that we have the reduced shadow automaton, we can 
explain how one can perform static analysis by \emph{abstract
  interpretation}~\cite{systematic} on concurrent systems, in an economic
way. The systematic design and proof of correctness of such abstract analysis is
left for a future article, the aim of this section is to give an intuition why
the computations of Section~\ref{programswithloops} are relevant to static
analysis by abstract interpretation. The idea is to associate, to each node $n$
of the shadow automaton, a set of values $A_n$ that program
variables can take if computation follows a transition path
whose last vertex is~$n$. Among the actions the program can take along this
scheduling, we consider only the \emph{greedy} ones, that is the ones which
execute all possible actions permitted by the dihomotopy class of schedulings
ending by $n$.

Suppose that we want to analyze the program
\begin{equation}
  \label{eq:ai-p}
  p^*
  \hspace{2ex}=\hspace{2ex}
  \Big(\P a.\pa{a:=a-1}.\V a\Big)^* \Big|\Big(\P a.\Big(a:=\frac{a}{2}\Big).\V a\Big)^*
\end{equation}
What are the possible sets of values reached, for $a$, starting with $a \in
[0,1]$? The associated shadow automaton~$S_p$ has been determined in
Example~\ref{ex:sh-det} (this automaton is reduced) together with relations,
that we will not be using in this article, yet. In many ways, this reduced
shadow automaton plays the role of a compact \emph{control flow graph} for
the program we are analyzing.
Calling 
$M_0=\includegraphics[scale=0.15]{sa_m1}$ and
$M_1=\includegraphics[scale=0.15]{sa_m0}$, $X_{M_0}$ has the effect on environment: $a:=a/2$ and
$X_{M_1}$ has as effect: $a:=a-1$.

We are now in a position to interpret the arrows of the shadow automaton as
simple \emph{abstract transfer functions} and produce a system of equations for
which we want to determine a least-fixed point, to get the invariant of the
program at the (multi-)control point which is the pair of the heads of the loops
of each process.
The interpretation on the shadow automaton now gives (ignoring the initial state $I$ in that picture, for
simplicity's sake) can be graphically pictured as:
\[
\vxym{
  0 \ar@(dl,ul)^{[a := a-1]}\ar@/^/[rr]|{[a := \frac a2]}&&\ar@/^/[ll]|{[a := a-1]}1\ar@(dr,ur)_{[a := \frac a2]}
}
\]
Given the abstract transfer functions on each edge of the shadow automaton,
we produce as customary the abstract semantic equations, one per node, by
joining all transfer functions correspond to ingoing edges to that node:
\begin{equation}
  \label{eq:ai-fp}
  \left(
    \begin{matrix}
      A_0\\A_1
    \end{matrix}
  \right)
  \qeq
  F
  \left(
    \begin{matrix}
      A_0\\A_1
    \end{matrix}
  \right)
  \qeq
  \left(
    \begin{matrix}
      I \cup (A_0-1)\cup (A_1-1)\\
      I \cup \frac{A_1}{2}\cup \frac{A_0}{2}
    \end{matrix}
  \right)
\end{equation}
This set of semantic equations can be seen as a least-fixed point equation, that
we can solve using any of our favorite tool, for instance Kleene iteration and
widening/narrowing, on any abstract domain, such as the domain of intervals as
in the example below.
The least-fixed point formulation that we are looking for is thus $A^{\infty}=
\bigvee_{[0,1]} F$, where~$F$ is the function defined in~\eqref{eq:ai-fp} and
$I=[0,1]$.
A Kleene iteration on this monotonic function~$F$ on the lattice of intervals
over~$\R$ reveals that $A_0^{\infty}=A_1^{\infty}=]-\infty,1]$.

We have presented this example in order to show how the reduced shadow automaton
can be used in order to use usual static analysis methods on concurrent
programs, avoiding state-space explosion as much as possible. It has the
advantage of being short, however it does not really show the main interest of
our technique: the scheduling automaton allows us to take in account properties
which tightly depend on the way the synchronizations constraint the executions
of the programs.

\section{Conclusion and Future work}
\label{concl}
We have presented an algorithm in order to compute a finite presentation of the
trace space of concurrent programs, which may contain loops. An application to
abstract interpretation has also described but remains to be implemented. In
order to give a simple presentation of the algorithm, we have restricted
ourselves here to programs of a simple form (in particular, we have omitted
non-determinism). We shall extend our algorithm to more realistic programming
languages in a subsequent article. Our approach can also be applied to languages
with other synchronization primitives (monitors, send/recv, etc.), for which
there are simple geometric semantics available.  There are also many possible
general improvements of the algorithm; the most appealing one would perhaps be
to find a way to have a more modular way of computing the total schedulings by
combining locally computed schedulings in~$\tspace{X}(x,y)$ with varying
endpoints~$x$ and~$y$. In a near future, the schedulings provided by the
algorithm will be used by our tool ALCOOL to analyze concurrent programs using
abstract interpretation, thus providing one of the first tools able to do such a
static analysis on concurrent programs without forgetting most of the possible
synchronizations during their execution.

On the theoretical side, we envisage to study in details and use the structure
of the index poset~$\C(X)$ which contains much more information than only the
schedulings of the program. Namely, it can be equipped with a structure of
\emph{prodsimplicial set} \cite{kozlov} (a structure similar to simplicial sets
but whose elements are products of simplexes), whose geometric realization
provides a topological space which is homotopy equivalent to the trace
space~$\tspace{X}$~\cite{raussen2010simplicial}. This essentially means
that~$\C(X)$ contains all the geometry of the trace space and we plan to try to
benefit from all the information it provides about the possible computations of
a program. Our ALCOOL prototype actually implements this computation -- using a
combinatorial presentation of the prodsimplicial sets known as \emph{simploidal
  sets} \cite{simploidal} -- which will be reported elsewhere.








\bibliographystyle{plain}
\bibliography{papers}
\end{document}